%% file: 1_main.tex
\documentclass{aastex631}
\usepackage{graphicx}
\usepackage{booktabs}
\usepackage{multirow}
\usepackage{float}
\usepackage{colortbl}
\usepackage{epstopdf} 
\usepackage{hyperref} 
\usepackage{xspace}
\usepackage{placeins}
\newcommand{\swiftj}{
\xspace}
\newcommand{\jj}{J151857\xspace}

\begin{document}

\title{Exploring the spin dependence on mass inclination and distance for the newly discovered black hole X-ray binary \swiftj}
\author[0009-0001-0703-2000]{Yujia Song}
\affiliation{Key Laboratory for Computational Astrophysics, National Astronomical Observatories, Chinese Academy of Sciences, Datun Road A20, Beijing 100012, China}
\affiliation{School of Astronomy and Space Sciences, University of Chinese Academy of Sciences, Datun Road A20, Beijing 100049, China}
\affiliation{Harvard-Smithsonian Center for Astrophysics, 60 Garden Street, Cambridge, MA 02138, USA}
\author[0000-0002-5872-6061]{James F. Steiner$^{\dag}$}
\affiliation{Harvard-Smithsonian Center for Astrophysics, 60 Garden Street, Cambridge, MA 02138, USA}
\author[0009-0001-0703-2000]{Tong Zhao}
\affiliation{Key Laboratory for Computational Astrophysics, National Astronomical Observatories, Chinese Academy of Sciences, Datun Road A20, Beijing 100012, China}
\affiliation{School of Astronomy and Space Sciences, University of Chinese Academy of Sciences, Datun Road A20, Beijing 100049, China}
\author[0000-0002-2268-9318]{Yuexin Zhang}
\affiliation{Harvard-Smithsonian Center for Astrophysics, 60 Garden Street, Cambridge, MA 02138, USA}
\affiliation{Kapteyn Astronomical Institute, University of Groningen, P.O.\ BOX 800, 9700 AV Groningen, The Netherlands}
\author[0009-0009-2549-1161]{Ningyue Fan}
\affiliation{Center for Astronomy and Astrophysics, Center for Field Theory and Particle Physics, and Department of Physics, Fudan University, Shanghai 200438, China}
\author[0000-0001-8670-4575]{Ole K\"{o}nig}
\affiliation{Harvard-Smithsonian Center for Astrophysics, 60 Garden Street, Cambridge, MA 02138, USA}
\author[0000-0002-6159-5883]{Santiago Ubach}
\affiliation{Harvard-Smithsonian Center for Astrophysics, 60 Garden Street, Cambridge, MA 02138, USA}
\affiliation{Departament de F\'{i}sica \& CERES-IEEC, Universitat Aut\'{o}noma de Barcelona, Bellaterra, Spain}
\author[0000-0001-6395-2066]{Josephine Wong}
\affiliation{Harvard-Smithsonian Center for Astrophysics, 60 Garden Street, Cambridge, MA 02138, USA}
\affiliation{Department of Physics and Kavli Institute for Particle Astrophysics and Cosmology, Stanford University, Stanford, CA 94305, USA}
\author[0000-0003-3057-5860]{Lijun Gou$^{\dag}$}
\affiliation{Key Laboratory for Computational Astrophysics, National Astronomical Observatories, Chinese Academy of Sciences, Datun Road A20, Beijing 100012, China}
\affiliation{School of Astronomy and Space Sciences, University of Chinese Academy of Sciences, Datun Road A20, Beijing 100049, China}
\author[0000-0003-3828-2448]{Javier A. Garcia}
\affiliation{Cahill Center for Astronomy and Astrophysics, California Institute of Technology, 1200 E. California Blvd., MC 290-17, Pasadena, CA 91125, USA}
\affiliation{X-ray Astrophysics Laboratory, NASA Goddard Space Flight Center, Greenbelt, MD 20771, USA}

\begin{abstract}

The black hole X-ray binary (BHXRB) \swiftj was discovered during its first outburst in March 2024. We review the archive of {\em NICER} observations from this outburst, focusing on the soft states. We select spectra for which the disk to total flux ratio exceeds 0.8 and the coronal scattering fraction $f_{\rm sc}$  is less than 25\%, conditions under which the accretion disk is expected to extend to the innermost stable circular orbit (ISCO) and remains geometrically thin. Through a continuum fitting analysis, we explore the dependence of spin on mass, inclination, and distance. We constrain the spin within the parameter space typical of  stellar-mass black holes (sBHs), assuming a mass around 10 solar masses, inclination angles between 20$^{\circ}$ and 60$^{\circ}$, and distances between 4 and 16 kpc. For fiducial parameters: a mass of 10 solar masses, a distance of 10 kpc, and an inclination angle of 40$^\circ$, a moderate spin of approximately 0.7 is obtained. However, precise determination of the spin will require accurate measurements of these parameters. Our analysis provides a framework to infer the spin and estimate its uncertainties once more precise measurements of mass, distance, and inclination become available.  As we demonstrate, lower inclination angles, greater distances, or larger black hole masses result in higher spin values. 

\end{abstract}

\keywords{accretion, accretion disks – black hole physics – X-rays: binaries}

\section{INTRODUCTION} 
\label{sec1}

To date, more than seventy stellar-mass black holes (sBHs) have been identified within the Milky Way \citep{remillardXRayPropertiesBlackHole2006, corral-santanaBlackCATCatalogueStellarmass2016a}. Almost all of them are located in X-ray binaries (XRBs), among the earliest X-ray sources discovered outside the Solar System. Recently, some additional black hole candidates have been identified through astrometric measurements, among them Gaia BH1, Gaia BH2, and Gaia BH3 \citep{el2023sun, el2023red, panuzzo2024discovery}. In early exploration of the X-ray sky, Sco X-1 and Cyg X-1 were particularly pivotal discoveries; while the former is a neutron star X-ray binary (NSXRB), the latter was the first confirmed black hole X-ray binary (BHXRB) \citep{giacconi1962evidence, oroszMASSBLACKHOLE2011a, reidTRIGONOMETRICPARALLAXCYGNUS2011a}. Most BHXRBs are classified as X-ray transients \citep{draghisSystematicallyRevisitingAll2024}, with only a small number exhibiting persistent emission \citep{gouEXTREMESPINBLACK2011a, steinerLOWSPINBLACKHOLE2014a}. Transients remain active for several months to years before fading into quiescence, whereas persistent sources exhibit sustained activity over decades or longer. 

The outburst cycle of a transient BH begins as it emerges from quiescence and typically progresses through a sequence of a low-hard state (LHS), intermediate state (IMS), and high-soft state (HSS), eventually returning to quiescence in a reversed-sequence of those states. The LHS is characterized by dominant power-law emission in hard X-rays, often with significant X-ray variability and featuring pronounced radio emission associated with a compact jet.  The X-ray variability includes quasi-periodic oscillations (QPOs), particularly the subclasses of low-frequency QPOs (LFQPOs) \citep{ingram2019review}.  High-frequency QPOs (HFQPOs) above roughly 20\,Hz are rarely detected, but when they are present, appear in the IMS which has spectral properties in-between the LHS and HSS.  In contrast to the LHS, the HSS is dominated by thermal multi-temperature disk radiation at around 1 keV, with weak or quenched radio jet activity.  

Currently, the primary spectral methods for determining sBH spins are continuum fitting \citep{zhangBlackHoleSpin1997a, mcclintock2014black, kulkarni2011measuring} and reflection spectroscopy \citep{fabianXrayFluorescenceInner1989}. Spin is a crucial property of a BH, as the no-hair theorem states that spin and mass fully characterize an astrophysical BH in general relativity. The continuum-fitting method assumes that the inner disk radius extends down to the innermost stable circular orbit (ISCO), while the reflection spectroscopy approach, though not requiring this assumption, faces challenges in simultaneously determining both the black hole spin and the inner disk radius in practical fitting\citep{bardeenRotatingBlackHoles1972c}. In the continuum-fitting method, the inner radius of the accretion disk ($R_{\rm in}$) is inferred by modeling the thermal emission from the disk, under the assumption of the relativistic thin disk model developed by \citep{shakura1973black, novikov1973astrophysics}. Observations suggest that after the system enters the HSS, the inner radius of the accretion disk reaches the ISCO and remains constant throughout this phase (e.g., \citealp{steinerCONSTANTINNERDISKRADIUS2010b, tanakaLewin1995, doneReview2007}). Assuming that the disk extends down to the ISCO, one can estimate the black hole spin by exploiting the one-to-one correspondence between the ISCO radius and the spin parameter. The continuum-fitting method requires independent measurements of several system parameters, including the black hole mass ($M$), the distance to the source ($D$), and the inclination angle ($i$) of the innermost accretion disk. It is commonly assumed that the black-hole spin angular momentum and the inner disk are aligned with the orbital angular momentum, i.e., that the black hole spin is aligned with the binary orbit. 

\swiftj (hereafter \jj) is a galactic transient source that was at first misclassified as a gamma-ray burst (GRB 240303A) and subsequently recognized as an X-ray transient \citep{lien2024grb}. The source was first localized by {\em Swift/BAT} at RA = 229$^\circ$.748, Dec = -57$^\circ$.375 with an onboard-calculated positional uncertainty of 3 arcminutes (90\% confidence radius). Subsequently, an immediate onboard localization by {\em XRT} refined the position to RA = 229$^\circ$.7375, Dec = -57$^\circ$.3633, with an uncertainty of 4.8 arcseconds, which lies within the {\em BAT} error circle. Spectral analysis of {\em Swift/XRT} data revealed an absorption column density of $N_{\rm H}=(5.60\pm0.06)\times10^{22}\,{\rm cm}^{-2}$ \citep{kennea2024swift}. This value exceeds the typical range of $N_{\rm H}$ for the Galactic sample, and is thus at the upper end of the distribution \citep{bekhti2016hi4pi}. Subsequent {\em Swift/XRT} observations detected a spectral softening trend, consistent with the characteristics of Galactic black hole X-ray binaries (BHXRBs) transitioning from the LHS to the HSS. This rapid transition was further confirmed by a bright radio flare detected with the {\em Australia Telescope Compact Array (ATCA}; \citealp{del2024swift}). The source’s distance was constrained to 4.48--15.64\,kpc using HI absorption measurements \citep{burridge2024hi}. Following the initial X-ray trigger, \citet{cowie2024discovery} conducted a 15-minute follow-up observation with {\em MeerKAT} at 1.28 GHz, confirming the presence of a radio counterpart. The observed spectrum and flux were found to align with that of an X-ray binary in the hard state, indicating that the source is either a black hole or a radio-bright neutron star X-ray binary. Optical and near-infrared (NIR; H-band) observations were conducted using the 60\,cm {\em Robotic Eye Mount (REM)} telescope in La Silla, Chile. The source was not detected in the optical band, likely due to significant dust extinction. However, a clear NIR counterpart was detected at the given coordinates, with an offset of approximately 2'', within the {\em XRT} error margin. The peak H-band magnitudes measured during the observation was $\rm H=12.33\pm0.06$ \citep{baglio2024near}. The transient brightening in the NIR  is consistent with the typical behavior of BHXRB transient outbursts during the initial rising phase. 

In this work, we investigate {\em Neutron star Interior Composition Explorer}  ({\em NICER}) X-ray data \citep{gendreau2012neutron} obtained during \jj's outburst and apply the X-ray continuum-fitting method to study its spin.   The first constraint on the dimensionless spin parameter of \jj was reported by \citet{pengInsightHXMTNICERNuSTAR2024} through a joint reflection spectral analysis of simultaneous {\em Insight-HXMT},{\em  NICER}, and {\em NuSTAR} observations. They found a moderate spin $a_* = 0.84^{+0.17}_{-0.26}$ and constrained the system's inclination angle of $21.1^{+4.5}_{-3.6}$ degrees. They derived a source distance of $5.8\pm2.5$ kpc by applying an empirical relation between the outburst profile and intrinsic power output. Subsequently, leveraging this distance and the relationship between the {\tt\string diskbb} normalization and the ISCO radius, they constrained the black hole's mass range $3.67 \pm 1.79$--$8.07 \pm 4.20\,M_{\odot}$.

\citet{mondal2024first} performed polarization and spectral analysis of this source using data from {\em NuSTAR} and the {\em Imaging X-ray Polarimetry Explorer} ({\em IXPE}). Their analysis of the {\em IXPE} data suggested a significant X-ray polarization detection from this source with a reported polarization degree (PD) of $1.34\pm0.27\%$ and a polarization angle (PA) of $-13^{\circ}.7\pm5^{\circ}.9$ using a model-independent approach, while the model-dependent analysis gave a PD of $1.18\pm0.23\%$ and a PA of $-14^{\circ}.01 \pm 5^{\circ}.80$.
However, the same {\em IXPE} data were independently analyzed by \citet{ling2024polarization}, who instead found PD$=0.3\%\pm0.3\%$ and inclination $=-24^{\circ}\pm26^{\circ}$, respectively, in the same 2--8\,keV band.  Notably, these results are statistically incompatible, with the \citet{ling2024polarization} results below the 99\% minimum degree of polarization confidence standard ($\rm{MDP}_{99}$) in all energy bins.  In particular, they obtain an upper limit on PD of 0.8\%. 

In addition to their polarimetric analysis, \citet{mondal2024first} performed a joint spectral study of the simultaneous {\em IXPE} and {\rm NuSTAR} observations, as well as independent {\em NuSTAR} observations in the HSS, exploring application of both the {\tt kerrbb} and {\tt JeTCAF} spectral models. For the two spectroscopic data sets which were fitted independently of one another, under the assumptions of a color-correction factor of 1.7 and a distance of 10 kpc, the {\tt kerrbb} model fits yielded estimated BH masses of $9.2\pm1.6$ and $10.1\pm1.7\,M_{\odot}$, with corresponding estimates for spin $0.6\pm0.1$ and $0.7\pm0.2$ and inclination of $35^{\circ}\pm7^{\circ}$ and $46^{\circ}\pm15^{\circ}$. The mass estimates from the {\tt JeTCAF} model, which does not include spin or inclination parameters, are found to be consistent with the values obtained using {\tt kerrbb}.
Note that, as we will show, the continuum-fitting spin values reported for {\tt kerrbb} are subject to significant systematic uncertainties, resulting in part from their assumption on both distance and use of a fixed (with luminosity) color-correction factor.
An overestimate of this factor forces a smaller inferred inner radius, thereby biasing the spin fit toward higher values, and vice versa.


In this paper, we employ a significantly more extensive set of {\em NICER} observations. Our sample comprises 61 spectra (compared to 19 in \citealp{pengInsightHXMTNICERNuSTAR2024}) with a total exposure of $\sim$100 ks (compared to $\sim$25 ks), enabling us to trace the source evolution to a lower luminosity of 2\% $L_{\rm Edd}$ and construct a more complete HID across a wider dynamic range (2--50\% $L_{\rm Edd}$, vs. 3--20\% in prior work). 

In the absence of dynamical parameter measurements for \jj, we adopt a conservative approach with agnostic and broad ranges explored for the system parameters, and present the spin's dependence on these parameters, systematically. Unlike the study by \citet{pengInsightHXMTNICERNuSTAR2024}, which relied on indirect inferences, our work directly determines the spin by systematically mapping a wide and agnostic parameter space.

The paper is organized as follows: The data and data reduction are described in Sect.~\ref{sec2}. The results of the spectral analysis are described in Sect.~\ref{sec3}. A discussion and a set of conclusions are presented in Sect.~\ref{sec4} and Sect.~\ref{sec5}, respectively.

\section{OBSERVATIONS AND DATA REDUCTION} 
\label{sec2}

The X-ray outburst of \jj was first observed by {\em NICER} on 2024 March 4 using the X-ray Timing Instrument (XTI). The observations continued until August 2024 and provided a wealth of data with $\sim 100$\,ks. {\em NICER} provides excellent capabilities for spectral studies in the soft X-ray energy range (0.2–-12 keV) due to its large effective area and high timing resolution, which makes it particularly suitable for characterizing low-temperature thermal emission components.
We use HEASOFT V6.33/NICERDAS v012 with CALDB release xti20240206 for the {\em NICER} data analysis. Initial data processing employed the standard pipeline tool {\tt\string nicerl2}\footnote{\url{https://heasarc.gsfc.nasa.gov/lheasoft/ftools/headas/nicerl2.html}}. We separately processed orbital night and day data using the {\tt\string threshfilter}. Since 2023 May 22, {\em NICER} has experienced a significant light leak, which has a minor effect on orbital night data but severely impacts orbital day observations, potentially leading to scientific data loss. For the orbit-day data, we set {\tt\string underonly\_range="0-500"} to extract the event files.  At the count rates encountered ($\lesssim 2000~{\rm counts\ s}^{-1}$ for the full array of 52 detectors), the data showed no significant pile-up effects, and deadtime losses were negligible. 

The spectra were extracted using {\tt\string nicerl3-spect}\footnote{\url{https://heasarc.gsfc.nasa.gov/lheasoft/ftools/headas/nicerl3-spect.html}}, with the ``{\tt\string scorpeon}" model prediction to estimate the background for the spectral analysis; we also set the minimum counts per bin to 20 using {\tt\string ftgrouppha}, while implementing the optimal binning method of \citet{kaastra2016optimal}. {\tt\string nicerl3-spect} also automatically applies a benchmark systematic error ($1.5\%$ across most of {\em NICER}'s range) using {\tt\string niphasyserr}. Due to high background levels in certain observations, the data were further screened by examining the individual good time intervals (GTIs). Observations with high background levels were subdivided into segments with a maximum interval of 200 seconds; segments exhibiting high background contamination were excluded from subsequent spectral analysis, thereby ensuring high signal-to-noise ratios. A total of 61 spectra were obtained via this processing, numbered N1--N61; these are presented in Table~\ref{tab:obs}.

\input{table_obs.tex}

\begin{figure}[htbp]
    \centering
    \includegraphics[width=\textwidth]{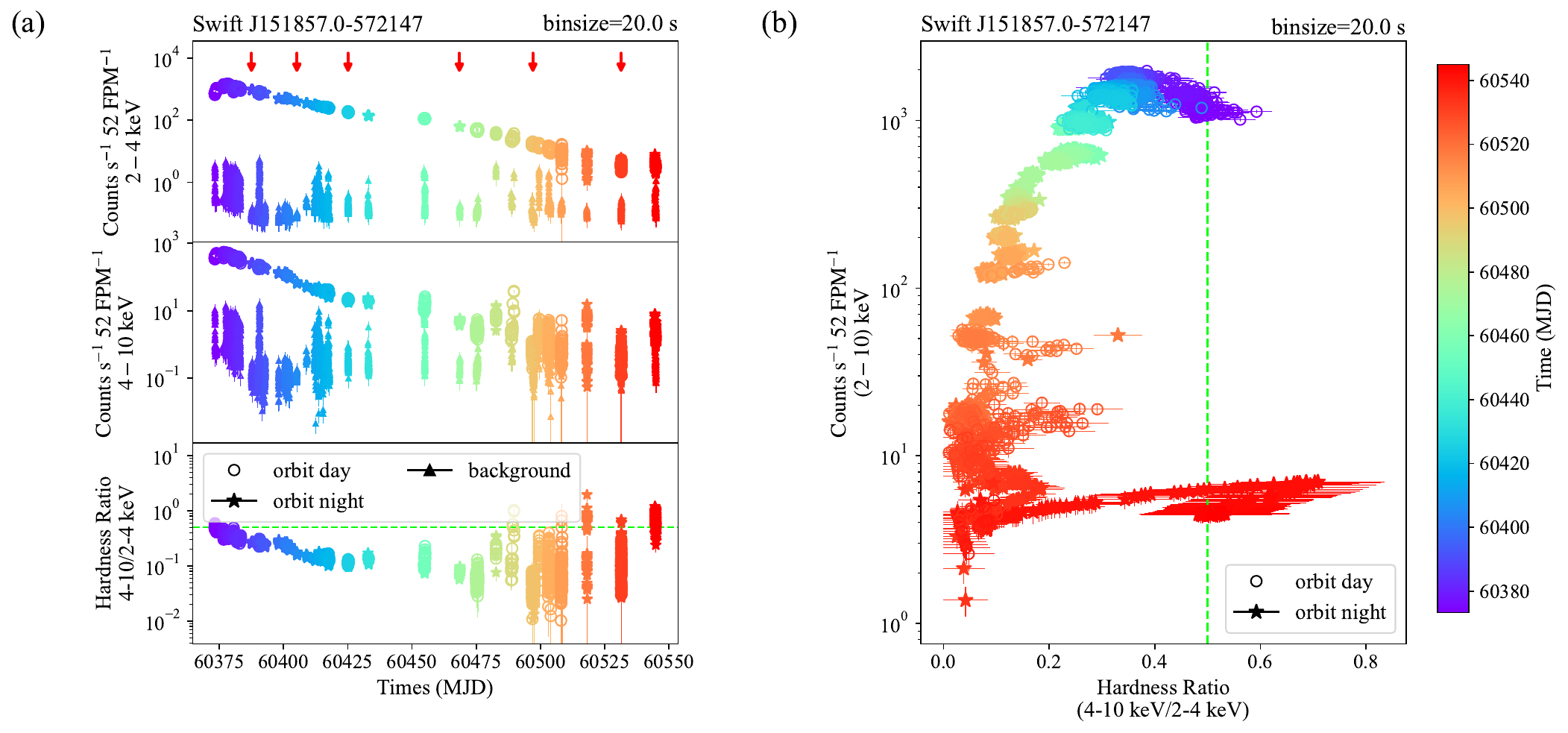}
    \caption{a) {\em NICER/XTI} light curves for 2024 outburst of \jj. The top and middle panels display the count rates in the 2–-4 keV and 4–-10 keV energy bands, respectively, with the background indicated below. The bottom panel presents the temporal evolution of the hardness ratio (defined as the 4--10 keV count rate divided by the 2--4 keV count rate). Red arrows indicate the six observations selected for the joint fit. b) HID corresponding to {\em NICER} observations shown in panel (a). Colors transitioning from purple (earlier times) to red (later times) depict the temporal sequence. The vertical line at a hardness ratio of 0.5 is included in the plot as a visual guide. The solid star and hollow circle symbols denote the orbit-night and orbit-day data, respectively. For clarity, data points with hardness uncertainties greater than 0.05 were smoothed by averaging the temporally adjacent 50 data points (i.e., generally spanning 1000 s).}
    \label{fig:HID}
\end{figure}

The light curves and corresponding background estimates were extracted separately using {\tt\string nicerl3-lc}\footnote{\url{https://heasarc.gsfc.nasa.gov/docs/software/lheasoft/ftools/headas/nicerl3-lc.html}} for the 2--4 keV and 4--10 keV energy bands, with orbit night and orbit day data processed independently. This distinct processing is crucial because the orbit day period, when {\em NICER} is exposed to sunlight, induces significantly elevated and spectrally distinct background noise, particularly at low energies. Fig.~\ref{fig:HID} displays {\em NICER}’s monitoring of \jj, presented in 20-second intervals. Panel (a) shows the source’s X-ray intensity evolution, while panel (b) presents the corresponding hardness-intensity diagram (HID), exhibiting the characteristic ``q"-shaped evolution common in BHXRB outbursts. 

\section{SPECTRAL ANALYSIS \& RESULT} 
\label{sec3}

In this section, we present the best-fit results of different spectral models. $\chi^2$-statistics were used for analyzing the spectral fits, which allows for incorporation of the calibration systematics (from the default settings of {\em NICERDAS}) in the analysis. The spectral analysis was performed using {\tt\string XSPEC} version 12.14.0 \citep{arnaud1996xspec}. Unless otherwise noted, all uncertainties reported in this paper are given at the 90\% confidence level. We adopt the interstellar-medium abundances of \citet{wilms2000apj} and the atomic cross sections of \citet{verner1996atomic}.

\subsection{The non-relativistic spectral model}\label{subsec:nonrel}

We initially applied a phenomenological spectral model, {\tt\string tbfeo*(ezdiskbb+nthcomp)}, to all data sets to perform a preliminary analysis. In this model, {\tt\string tbfeo} \citep{wilms2000apj} accounts for Galactic hydrogen absorption and allows us to vary the abundances of oxygen and iron, {\tt\string ezdiskbb}\footnote{{\tt\string ezdiskbb} is a multiple blackbody disk model with zero-torque inner boundary. This model serves as an alternative to the standard {\\tt\string diskbb} model, incorporating a non-zero torque boundary condition at the inner disk edge.}  \citep{zimmerman2005multitemperature}  models the multicolor-blackbody emission from the accretion disk with zero torque inner boundary, and {\tt\string nthcomp} \citep{zdziarskiNTHCOMP1996} represents thermal Comptonization of the disk's thermal emission in a hot corona.

 \begin{figure}[htbp]
    \centering
        \includegraphics[width=\textwidth]{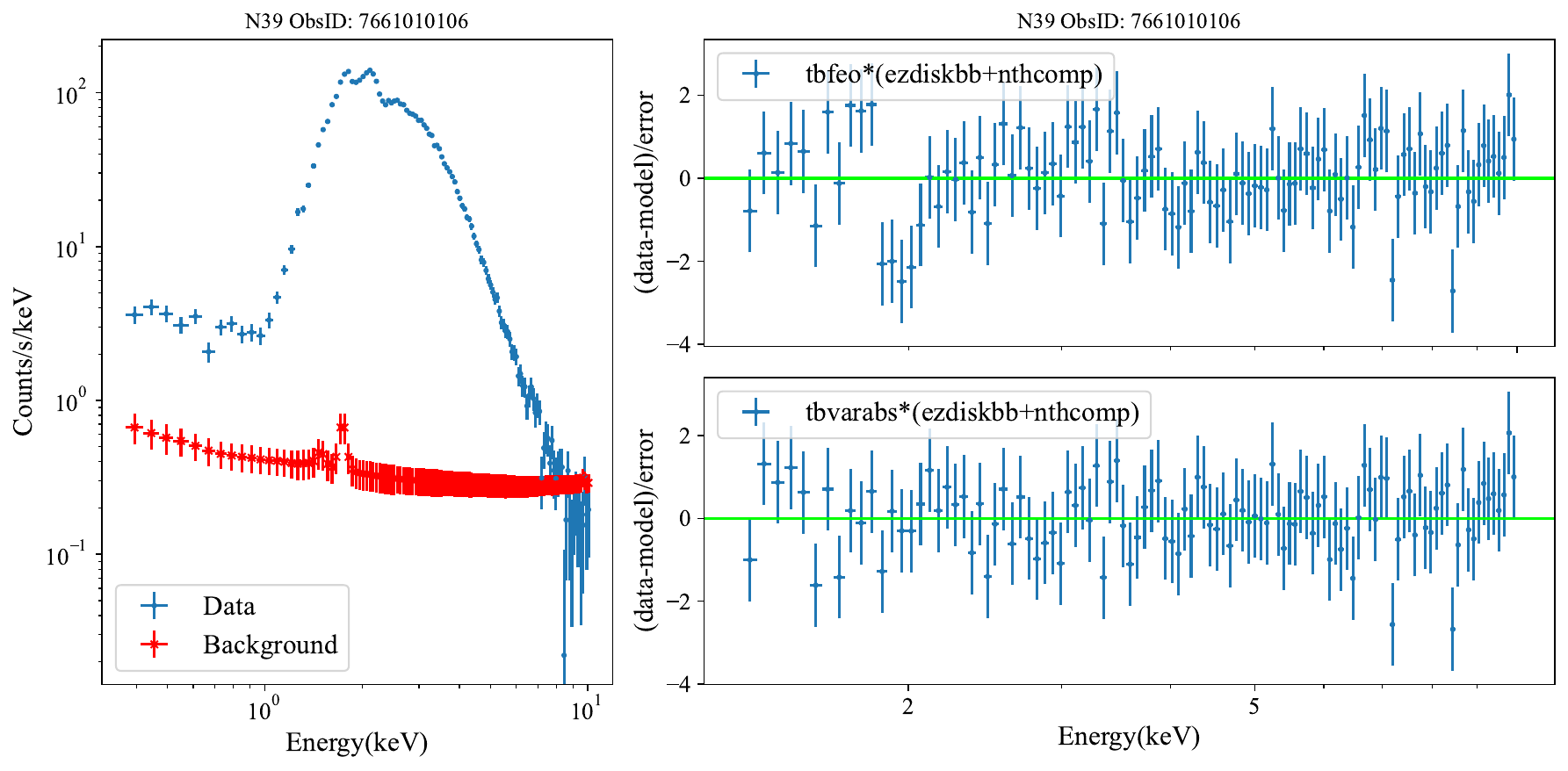}
        \label{fig:del}
    \caption{Illustration of spectral characteristics and fitting results for an orbit night spectrum (ObsID: 7661010106, N39) of \jj. Blue represents the signal, while red indicates the background. Left: Spectrum in the 0.4-10 keV range. A significant flat signal below 1 keV is observed. Right: Spectral fitting results. Upper panel: Best-fit with {\tt\string tbfeo(ezdiskbb+nthcomp)} shows residuals around 1.3 keV and 1.84 keV. Lower panel: Improved fit with {\tt\string tbvarabs(ezdiskbb+nthcomp)} significantly reduces low-energy residuals.}
    \label{fig:energy}
\end{figure}

We initially explore the 0.4--10 keV energy range while applying a 1\% systematic error term to all models. This empirical practice accounts for both instrumental calibration uncertainties. As shown in Fig.~\ref{fig:energy}, we observe a flat signal below approximately 1\,keV, with the hydrogen column density derived from the {\tt\string tbfeo} being about $5\times 10^{22}$ cm$^{-2}$. This feature results from the low-energy shelf of the response. Noting that at its high $N_{\rm H}$, the source emission is attenuated by photoelectric absorption more than a 100-fold at energies below $\sim 1.25$~keV, we adopt a pragmatic range 1.25--10 keV in all subsequent analysis. This is in order to be highly sensitive to the source emission while unaffected by the low-energy shelf's calibration. We note that late in the outburst when the count rate was lower, in some spectra the background exceeded the signal at high energies. In those instances, we limited the high-energy range of our analysis to correspond to the approximate crossover of source-dominated to background-dominated bands.  In one of the more extreme examples, for N60, only the 1.25--6 keV band was used. We also identified spectral features around 1.3 keV and 1.84 keV (see Fig.~\ref{fig:energy}), likely originating from the K-edges of Mg and Si, respectively \citep{rolleston2000galactic,zeegers2017absorption,zeegers2019dust}. 

\begin{figure}[htbp]
    \centering
    \includegraphics[width=0.8\textwidth]{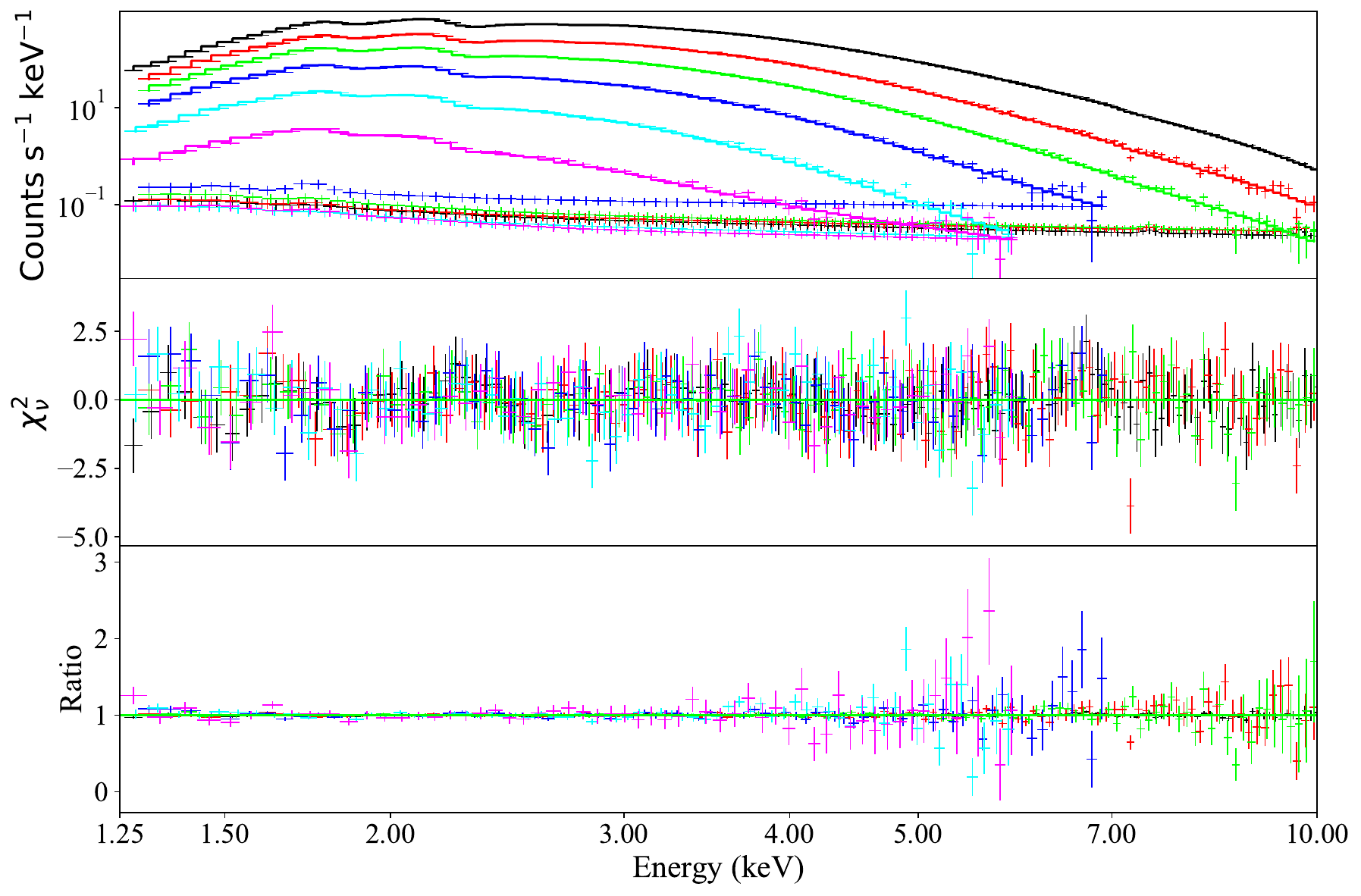}
    \caption{The joint fit results of Model 1 \texttt{tbvarabs*(ezdiskbb+nthcomp)} for the six spectra from different periods are presented. These include the background data, residuals, and ratio plots, respectively.}
    \label{fig:joint}
\end{figure}
\begin{figure}[htbp]
    \centering
    \includegraphics[width=0.8\textwidth]{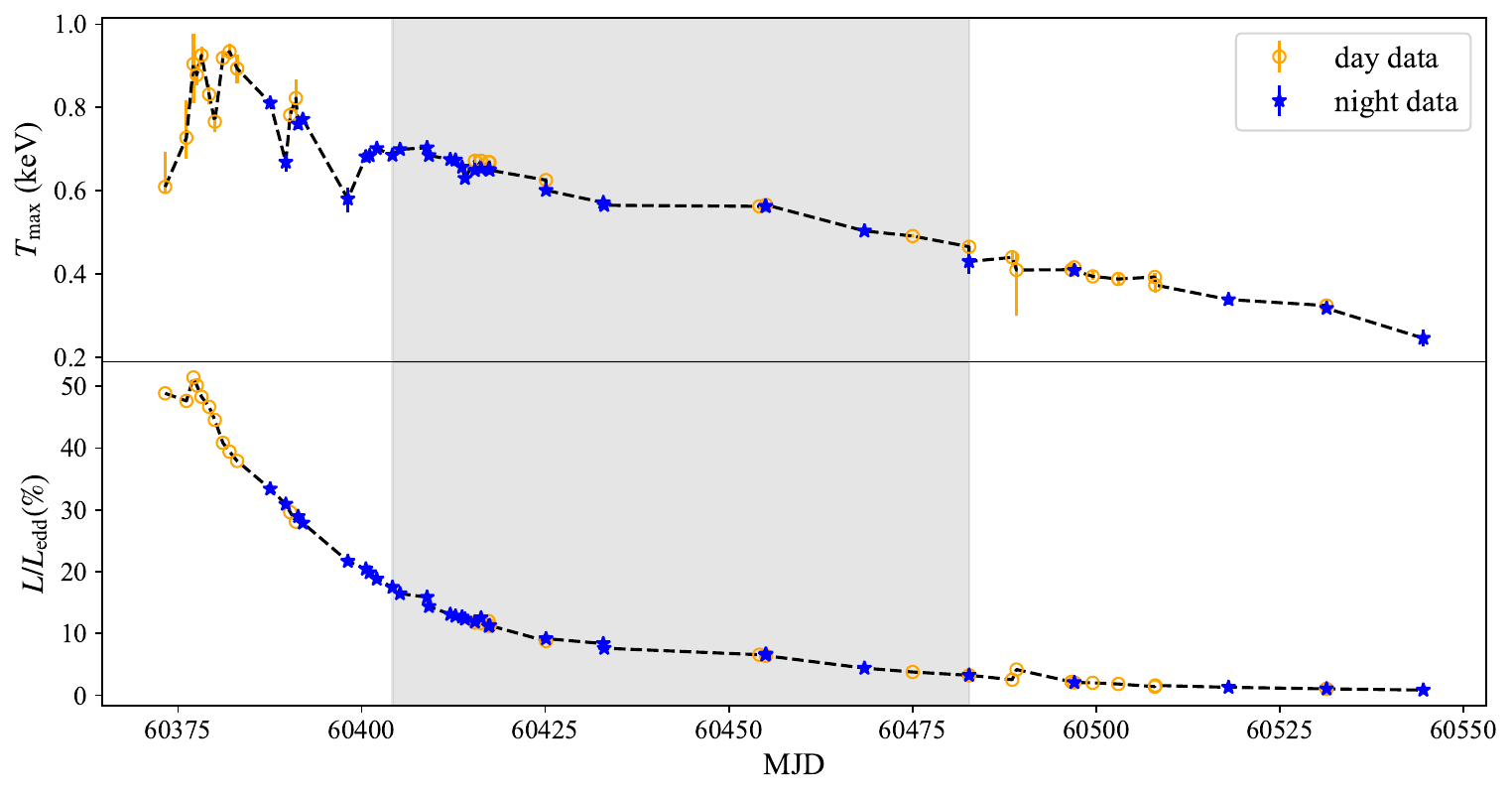} 
    \caption{Fitting results of model: {\tt\string tbfeo*(ezdiskbb+nthcomp)}. The disk temperature and the Eddington ratio throughout the outburst. The grey shaded regions represent data with an Eddington ratio between 3--30\% and a disk fraction greater than 0.8.}
    \label{fig:ezdisk}
\end{figure}

To account for these features, we replaced the {\tt\string tbfeo} with the more flexible {\tt\string tbvarabs}, allowing H, Fe, Mg, and Si column densities to vary freely while requiring that the relative abundance of Mg is equal to that of Fe (i.e., in Solar units). All other parameters were fixed to their default values. This adjustment resulted in a significantly improved fit. By visually inspecting the light curve, we selected six orbit night spectra (indicated by red arrows in Fig.~\ref{fig:HID}) from different periods of the outburst and spanning a range of luminosities and performed a joint fit using Model 1: {\tt\string tbvarabs*(ezdiskbb+nthcomp)}. The abundance and column-density parameters of the {\tt\string tbvarabs} for the different spectra were linked, while the other components were allowed to vary freely. The best-fitting column-density and abundances were $N_{\rm H}=(5.06^{+0.12}_{-0.12})\times10^{22}$~cm$^{-2}$,  Fe and Mg abundances $1.81^{+0.16}_{-0.15}$ times solar, and Si $2.29^{+0.11}_{-0.11}$ times solar. As for high photon indices, the spectrum drops off quickly due to the limited high-energy band, making it difficult to robustly constrain high values of the photon index $\Gamma$. Therefore, we bounded the fitted $\Gamma$ to the range between 1.3 and 3.5 \citep{esin1997advection,mcclintock2006spin}. In addition, the coronal electron temperature cannot be constrained with NICER alone, so we fixed it at a reasonable value of  $kT_{e} =$ 100 keV. The results of the joint fit are shown in Table~\ref{tab:joint} and Fig.~\ref{fig:joint}. It should be noted that cases of $\chi^2/\nu<1$ may arise from the lower signal-to-noise in the high-energy band. The parameters of {\tt\string tbvarabs} are fixed based on these joint-fit results in all subsequent fitting of the outburst data. Fig.~\ref{fig:ezdisk} shows the evolution of the fitted spectral parameters during the outburst.

\begin{figure}[htbp]
    \centering
    \includegraphics[width=0.9\textwidth]{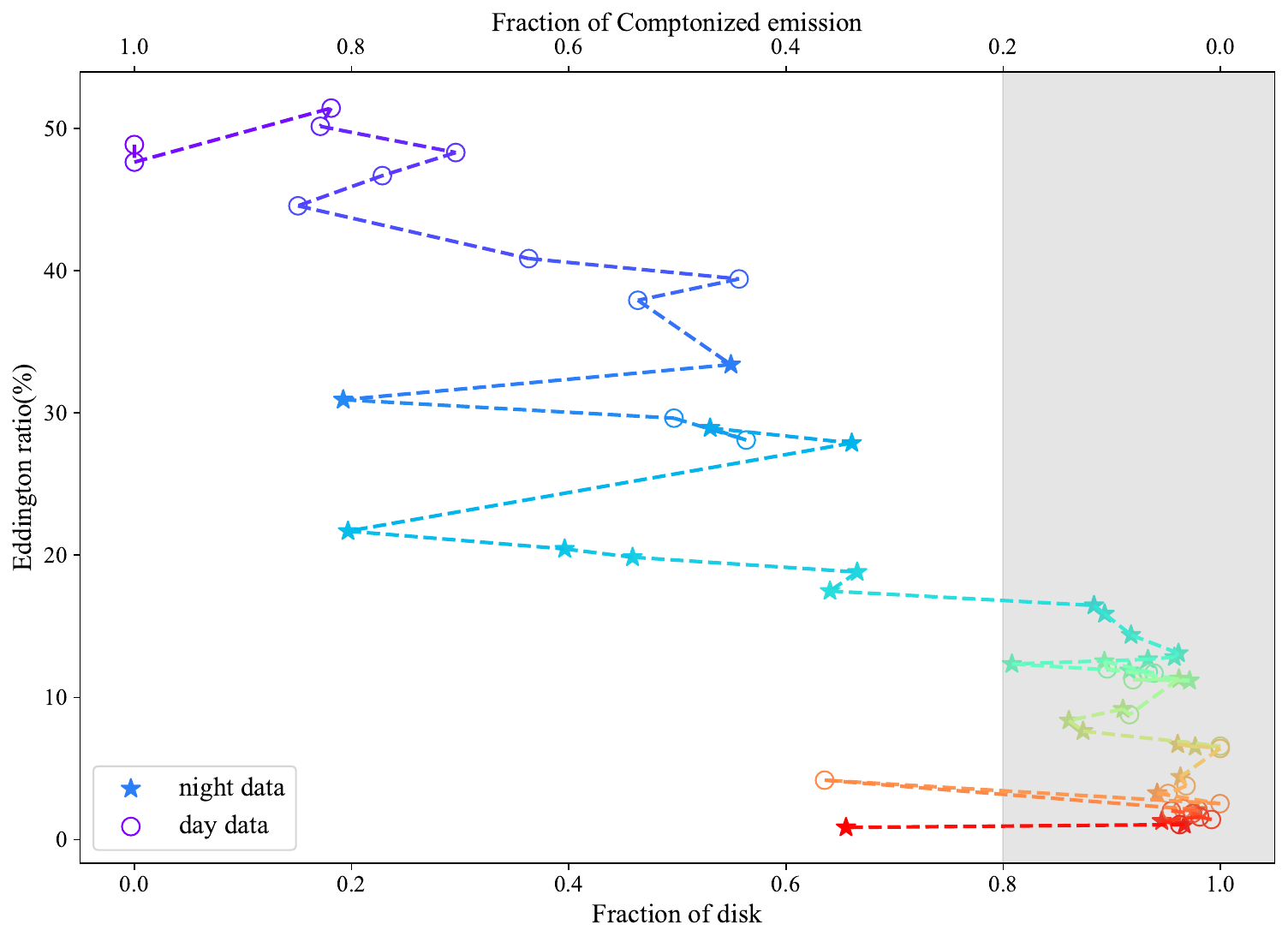}
    \caption{The percentage of flux contributions from different components, calculated using {\tt\string cflux}. The star shapes represent the orbital night data, while the hollow circular shapes represent the orbital day data. The time sequence follows the same color gradient as in Fig.~\ref{fig:HID}. The shaded region represents the region where the disk fraction is greater than 0.8, which is consistent with the shaded data in Table~\ref{tab:obs}.}
    \label{fig:L_evo}
\end{figure}

The {\tt\string energies}\footnote{\url{https://heasarc.gsfc.nasa.gov/xanadu/xspec/manual/node95.html}} command was used to define an extended energy grid. We calculated the flux from 0.01-100 keV for each spectrum using the {\tt\string flux}\footnote{\url{https://heasarc.gsfc.nasa.gov/xanadu/xspec/manual/node97.html}} command. Assuming a fiducial BH mass of $10\,M_{\odot}$ and a distance of 10 kpc, we obtained the unabsorbed luminosities for each observation. Additionally, we also used the {\tt\string cflux}\footnote{\url{https://heasarc.gsfc.nasa.gov/xanadu/xspec/manual/node292.html}} model to derive the flux contributions from the disk and power-law components separately, and to calculate the disk fraction within the 0.01–-100 keV energy band. The results are shown in Fig.~\ref{fig:L_evo}. Interestingly, \jj exhibits a flux decay of more than two orders of magnitude between the hard-to-soft and soft-to-hard transitions, a behavior previously reported only in a few systems, such as Swift J1727.8$-$1613 and V4641 Sgr \citep{podgorny2024recovery, revnivtsev2002super}. In the following analysis, only data with a luminosity between 3\% and 30\% $L_{\rm Eddington}$ and a disk fraction greater than 0.8 were selected for further investigation \citep{shakura1973black,reynolds2021observational}. These data are highlighted in Table~\ref{tab:obs} with shading. Under these conditions, the system is primarily dominated by the thermal emission from the optically thick and geometrically thin accretion disk, and simultaneously we have a grounded basis for the assumption that the disk radius extends to the ISCO (e.g., \citealp{liu2008precise, steiner2009measuring, gou2011extreme, belczynski2024common}).

\subsection{The relativistic spectral model}

Following our preliminary analysis, we selected the relativistic accretion disk model, {\tt\string kerrbb} for conducting further investigation. This model used ray tracing to derive the accretion disk's spectrum \citep{li2005multitemperature}. We fixed the hydrogen column density and abundances as described in the Model~1 fits from Sect.~\ref{subsec:nonrel}.
For Model~2, defined as {\tt\string tbvarabs(simplcut*kerrbb)}, we used {\tt\string simplcut}—an extension of the {\tt\string simpl} Comptonization model that incorporates an exponential cutoff in the power-law component \citep{steiner2009simple,steiner2017self}. We first explore a fiducial black hole mass of $10\,M_\odot$, a distance of 10 kpc, and an inclination angle of 40 degrees \citep{mondal2024first}. We focus in particular on $f_{\rm{sc}}$, the scattered fraction which describes the relative amplitude of the Comptonization component. 

\input{table_joint.tex}

As discussed for Model~1, owing to the lack of coverage above 10 keV and the intrinsically weak hard X-ray emission in the high/soft state (HSS), the empirical constraint on the photon index is often rather poor in this state. Therefore, for these soft-state spectra we restrict  $\Gamma$ to the range 2--3.5 (e.g., \citealp{remillardXRayPropertiesBlackHole2006}). For some spectra where the lower limit of $f_{\rm{sc}}$ could not be constrained, we bounded $f_{\rm{sc}}$ to values $>$ 0.01. The coronal electron temperature which gives rise to a hard X-ray turnover (outside {\em NICER}'s bandpass) in {\tt\string simplcut} was frozen at 100 keV and the reflection fraction was set to 0 which is equivalent to all reflection emission scattering elastically off the disk — acting like a mirror — e.g., when the disk is very highly ionized. The spectral fit results verify that the $f_{\rm{sc}}$ values are below 0.25, in-line with our previous selection criteria requiring a disk fraction greater than 0.8, i.e., suggesting that the system was in a disk-dominated state during such periods.  To implement the zero-torque boundary condition at the inner edge of the accretion disk, we set the parameter $\eta$ in the {\tt kerrbb} model to zero (but see \citealp{mummery2024continuum}). In addition, we accounted for the disk self-irradiation and limb-darkening effects by setting rflag = 1 and lflag = 1, respectively. The spectral hardening factor was fixed at 1.7 \citep{shimuraTakahara95}, a widely adopted value for black hole X-ray binaries, reflecting typical deviations from a pure blackbody due to radiative transfer effects in the disk atmosphere \citep{zhangBlackHoleSpin1997a}. Most of the spectra are well fitted although a few  show residual features near 2.2 keV, which are likely caused by calibration residuals related to the Au edge of the {\em NICER} instrument\footnote{\url{https://heasarc.gsfc.nasa.gov/docs/heasarc/caldb/nicer/docs/xti/NICER-Cal-Summit-ARF-2019.pdf}}.  Including a narrow Gaussian model to fit this feature effectively improves the fit and has minimal effect on other fit parameters. Given the robustness of the other parameters, we decided not to include this component in our analysis. The {\tt\string kerrbb} model gave a moderate spin, with a weighted average of $a_{*}=0.604\pm0.004$ and a standard deviation of $\sigma_{a*}=0.028$. 

\begin{figure}[htbp]
    \centering
    \includegraphics[ width=0.8\textwidth]{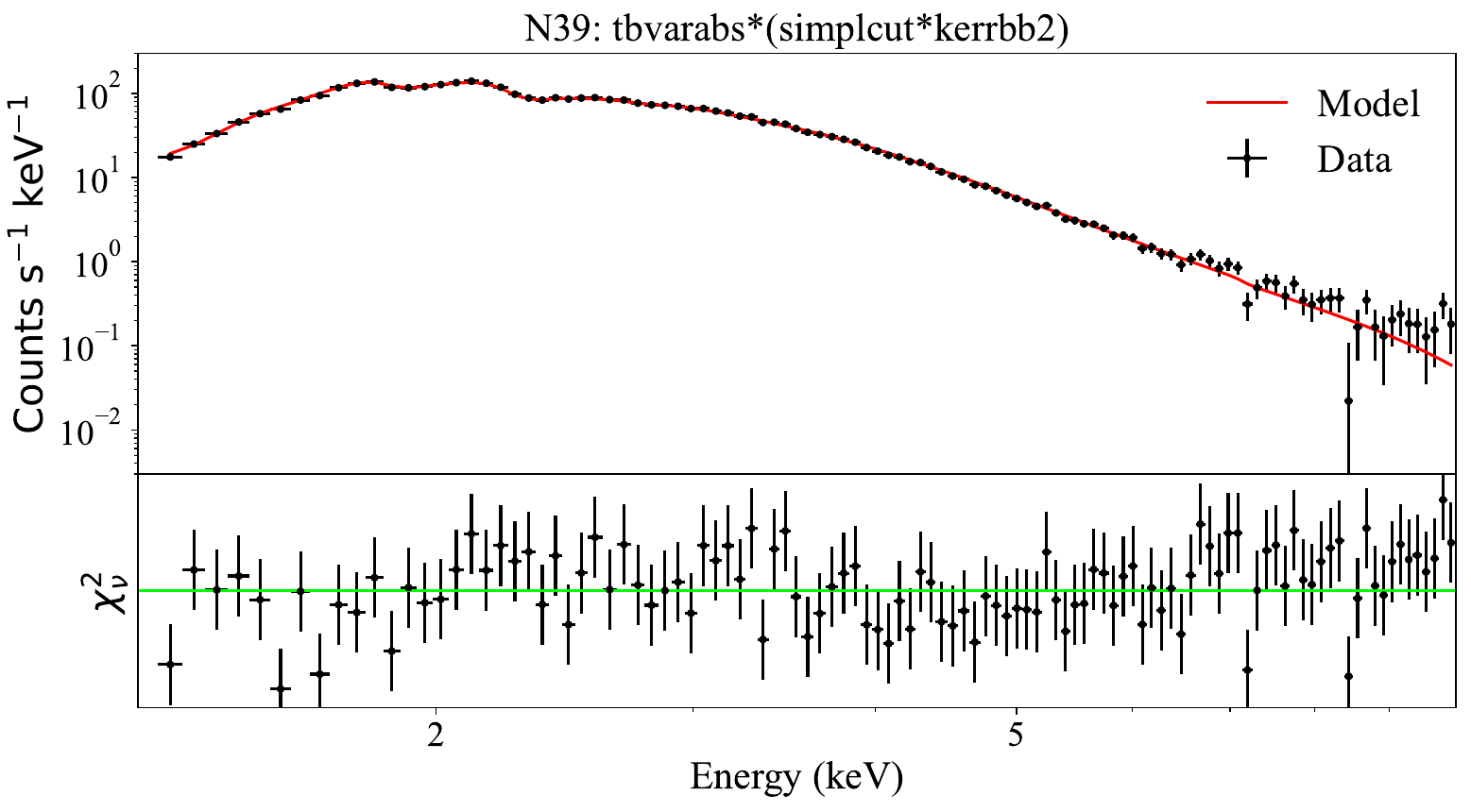}
    \caption{Spectral fits for N39 with Model 3: {\tt\string tbvarabs(simplcut*kerrbb2)}. Black dots represents the signal, while red line indicates the model. The variation in the residual plot is minimal.}
    \label{fig:N39}
\end{figure}

\begin{figure}[htbp]
    \centering
    \includegraphics[ width=0.8\textwidth]{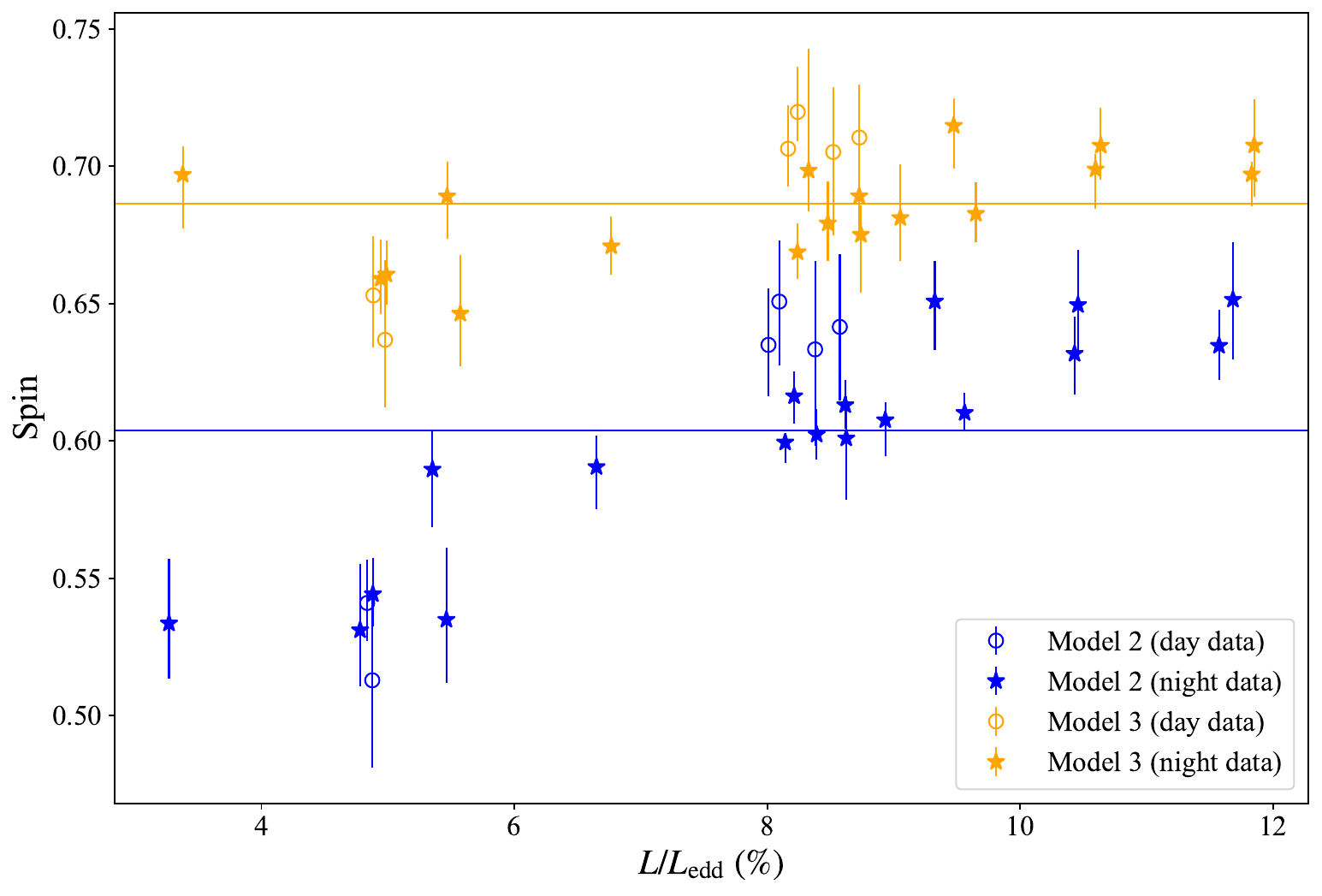}
    \caption{The relationship between black hole spin and luminosity predicted by Model~2: {\tt\string tbvarabs*(simplcut*kerrbb)} (blue) and Model~3: {\tt\string tbvarabs*(simplcut*kerrbb2)} (orange). Horizontal lines indicate the weighted mean spin values corresponding to each model.}
    \label{fig:model}
\end{figure}

We next replace the {\tt\string kerrbb} model with {\tt\string kerrbb2} \citep{mcclintock2006spin}. The {\tt\string kerrbb2} model combines the disk models {\tt\string bhspec} and {\tt\string kerrbb}, where {\tt\string bhspec} is used to determine the color-correction factor $f_{\rm{col}}=T_{col}/T_{eff}$ \citep{davis2005relativistic}. The complete Model 3 is: {\tt\string tbvarabs(simplcut*kerrbb2)}. First, in the manner  described in \citet{mcclintock2006spin}, we use {\tt\string bhspec} simulations fitted with kerrbb in order to generate a table of color correction factors. This f-table is then automatically read and applied by {\tt\string kerrbb2} as a function of the accretion rate $\dot{M}$ and spin $a_{*}$.  {\tt\string kerrbb2} takes into account the disk’s self-irradiation effects (rflag=1) and the effects of limb darkening (lflag=1). We again set the torque at the inner edge of the accretion disk to zero ($\eta=0$).  We have adopted an $\alpha$-viscosity parameter of 0.01 \citep{mcclintock2006spin}. Table \ref{tab:kerrbb2} summarizes the fitting results. As shown, with the exception of observations from the last two orbital days, all the luminosity ratios are in the range of 3--30\%. After excluding observations from the last two orbital days, the remaining data yield consistent, moderate spin values with a weighted average of $a_{*}=0.686\pm0.004$ and a standard deviation of $\sigma_{a*}=0.019$. In addition, we have calculated the Eddington-luminosity based on the fitting results obtained from  {\tt\string kerrbb2}. Compared to the results from the {\tt\string kerrbb} model with f=1.7, the $\chi^2$ value increased slightly by approximately 3\%, and the inferred spin value is now determined to be somewhat higher. We illustrate the fitting results for N39 (Fig.~\ref{fig:N39}) which shows its best fits using each model. We also present in Fig.~\ref{fig:model} the relationship between spin and luminosity as predicted by Model~2 and Model~3.

\subsection{Spin error analysis} \label{subsec:error}

\begin{figure}[htbp]
    \centering
    \includegraphics[width=\linewidth]{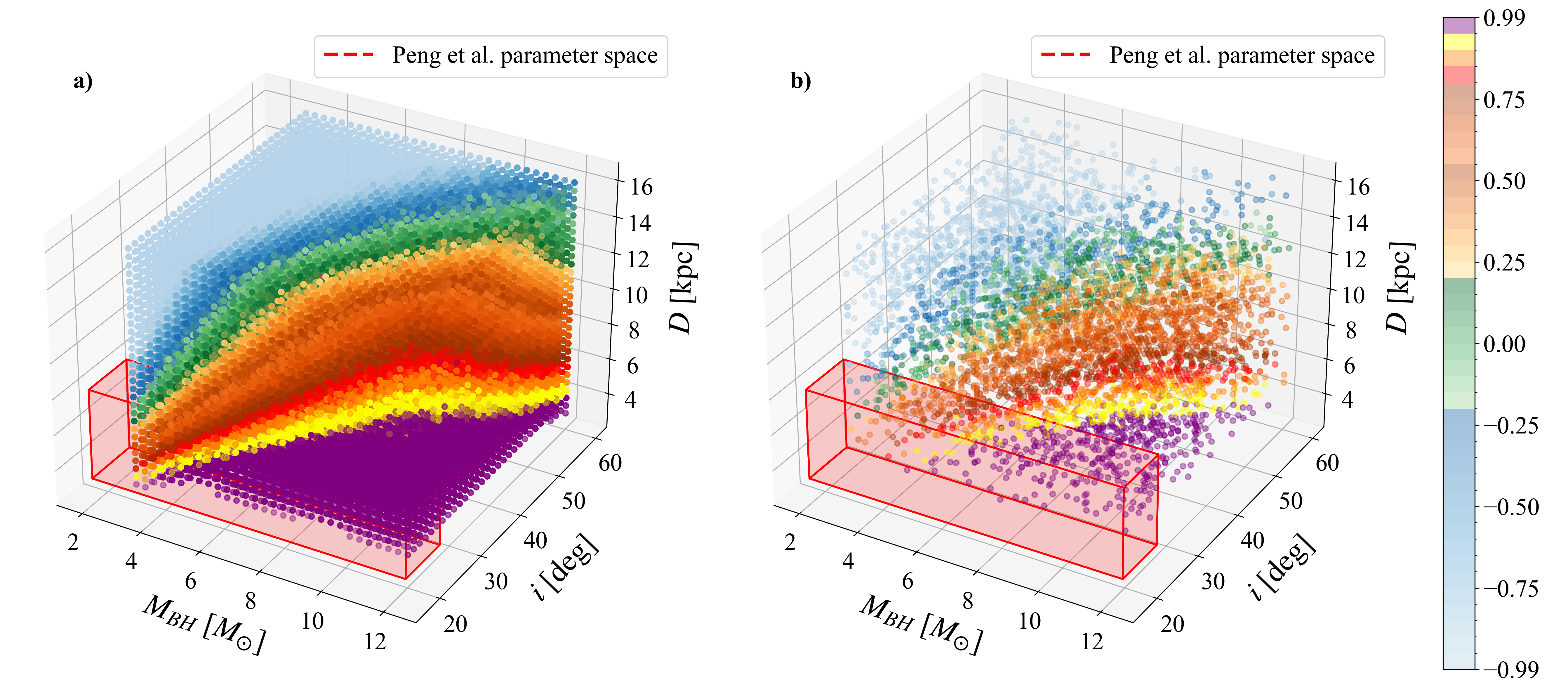}
    \caption{Spin as a function of the BH mass, inclination and distance with Model3:\texttt{tbvarabs*(simplcutx*kerrbb2)}. To ensure data quality, we excluded fitting results with $\chi^{2}/\nu>400$, regardless of the degrees of freedom. a) shows the spin scatter plot of 470,063 points derived from the final filtered fitting results. b) For enhanced visual clarity, we randomly selected 5,000 representative points to illustrate the spin distribution across the parameter space. The red wireframe cube delineates the parameter region derived by \citet{pengInsightHXMTNICERNuSTAR2024}}
    \label{fig:grid_xkbb}
\end{figure}

Since the system parameters have not been dynamically determined, we explore the spin dependence over a broad-range of plausible values for these parameters: $M=3-12\ M_{\odot}$, $i=20-60$ degrees, and $D=3-16$ kpc \citep{mondal2024first,burridge2024hi}. This mass range is in line with the observed distribution of black hole X-ray binary masses which is found to peak near $\sim 8\,M_{\odot}$ \citep{ozel2010black, farr2011mass}. We generated a $30\times30\times30$ linear grid spanning these ranges. Model 2: {\tt\string tbvarabs*(simplcut*kerrbb)} and Model 3: {\tt\string tbvarabs*(simplcut*kerrbb2)} were then applied to filter the data. We first selected the spectra with a disk fraction greater than 0.8 to ensure a thermally dominant state and then fitted corresponding data with these models \citep{steiner2009simple}. The spectra that satisfy the following criteria — an Eddington-scaled luminosity ratio between 3\% and 30\%, and a scattering fraction $f_{\rm{sc}}<0.25$ — were used to determine the spin \citep{steiner2009measuring}. Due to an optical leak in {\em NICER} since 22 May 2023\footnote{\url{https://heasarc.gsfc.nasa.gov/docs/nicer/analysis_threads/light-leak-overview/}}, orbital day data produce a higher undershoot rate and at sufficient rates have less-certain gain calibration.  Accordingly, we anchor our analysis on the orbital night data given the precise spectroscopy modeling at our focus. However, we note that fitting the corresponding orbital day data yields results comparable to those obtained from the orbital night data, using the fiducial system parameters (see Table~\ref{tab:kerrbb2}). We show a scatter diagram of spins in the parameter space in Fig.~\ref{fig:grid_xkbb}. We have also calculated and plotted the average of all spin values within each cubic grid cell. The absence of color for several cells, particularly near grid boundaries, is an indication that for those cells, no spectra within that parameter space satisfy the above conditions.

\input{table_xkbb}

\section{DISCUSSION} \label{sec4}

\subsection{The truncation radius in inner disk}\label{subsec:ezdiskbb}
\begin{figure}[htbp]
    \centering
    \includegraphics[width=0.9\textwidth]{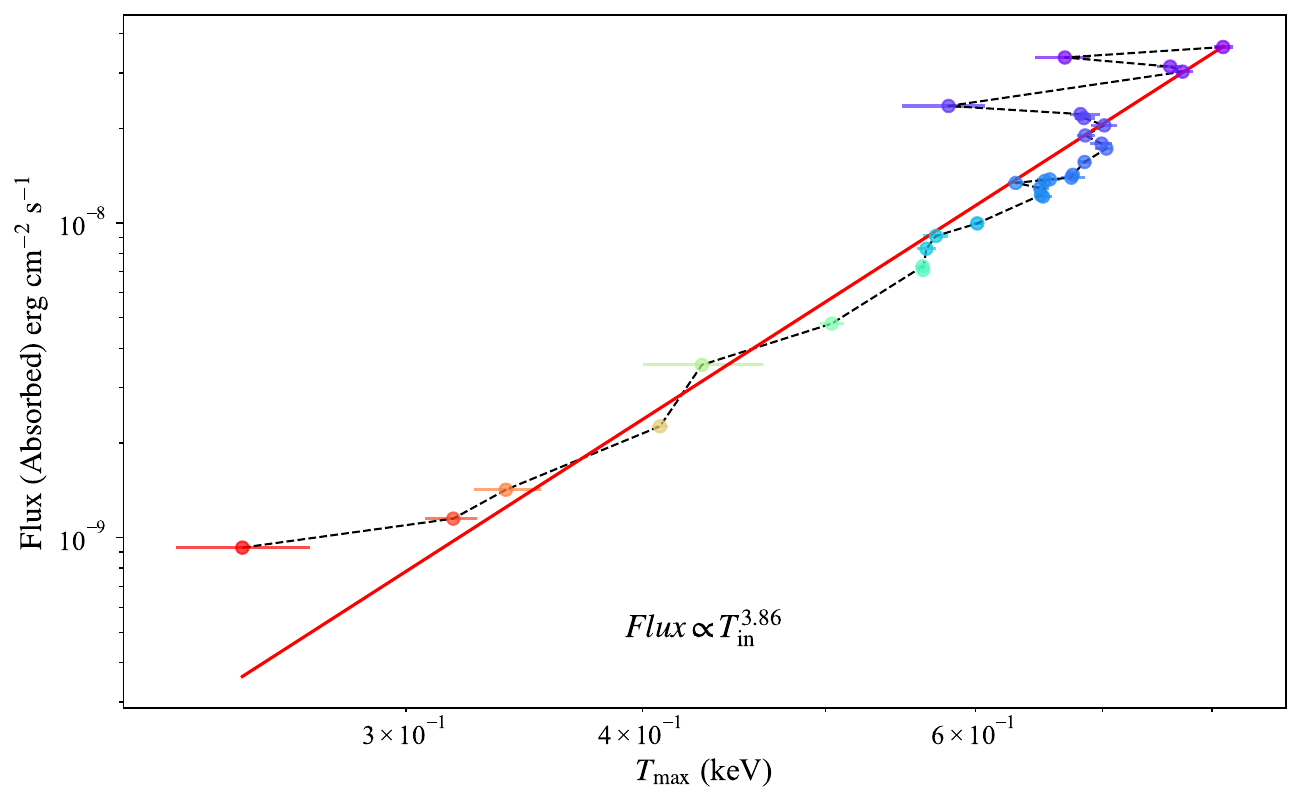}
    \caption{Accretion disk properties of \jj in the soft state. The relationship between color temperature from the spectral fit and the flux (0.01-100\,keV). The overlaid curve represents the best-fit power-law model.
}
    \label{fig:tin}
\end{figure}

In the model {\tt\string ezdiskbb}, $N_{\rm ezdiskbb}$ is the disk normalization: $(1/f^{4})(R_{\rm in}/D)^{2}{\rm cos}\;i$ \citep{zimmerman2005multitemperature}, where $R_{\rm in}$ is the apparent inner disk radius in units of km, $D$ is the distance to the source in units of 10 kpc, $i$ is the inclination, and $f_{\rm{col}}$ is the color to effective temperature ratio. Assuming a distance of 10 kpc \citep{burridge2024hi}, an inclination of 40 degrees \citep{mondal2024first}, and a fiducial color factor of 1.7 for this soft state black hole \citep{zhangBlackHoleSpin1997a}, the average inner radius of the orbit night spectra is very roughly estimated to be $R_{\rm in}\approx76$\,km, which would correspond to $\sim 2.2\,R_g$ for a $10\,M_\odot$ black hole. This estimate does not account for relativistic effects and would scale with different choices for mass, inclination, or distance. Our ISCO radius is slightly closer than $R_{\rm{in}}=2.6\pm0.04\,R_g$ reported by \citet{pengInsightHXMTNICERNuSTAR2024}, which was anchored on an empirical relationship and derived using the normalization of diskbb. We note that discrepancies are expected given the significant differences in the values for distance and inclination in this work compared to theirs.
The trend we observe between disk radius and temperature leads us to believe that $R_{\rm in}$ has reached the ISCO during this period, as has been commonly inferred for systems in the soft state. As a result of the zero-torque boundary condition adopted in the \texttt{ezdiskbb} model, the disk’s maximum temperature ($T_{\rm max}$) occurs slightly away from the innermost edge, rather than at the edge itself. The radial temperature profile of the accretion disk therefore differs from that assumed in the widely-used {\tt\string diskbb} model \citep{mitsuda1984}. Nevertheless, the observed relationship between flux and temperature (often approximated as $F \propto T^4$ for a standard disk) remains a useful diagnostic of the disk properties. In this context, the inner disk temperature ($T_{\rm in}$) is approximately related to the maximum disk temperature ($T_{\rm max}$) by $T_{\rm in} \approx 2.05 \times T_{\rm max}$, as derived from the temperature profile of the {\tt\string ezdiskbb} model \citep[Eq.~5 of][]{zimmerman2005multitemperature}. This relationship is shown in the Fig.~\ref{fig:tin}, we have fitted a power-law to the flux versus temperature data. We determine a scaling $\propto T_{\rm in}^{3.86
\pm0.49}$.  Therefore, the relationship observed between flux and  temperature suggests that we measured a constant radius (the ISCO) suitable for spin analysis. This empirical grounding to our approach is robustly demonstrated by the $\sim$9\% consistency of the inner-radius found by {\tt\string kerrbb2} across the thin-disk luminosity range.

\subsection{The influence of the photon index \texorpdfstring{$\Gamma$}{Gamma}}

\begin{figure}[htbp]
    \centering
    \includegraphics[width=0.9\textwidth]{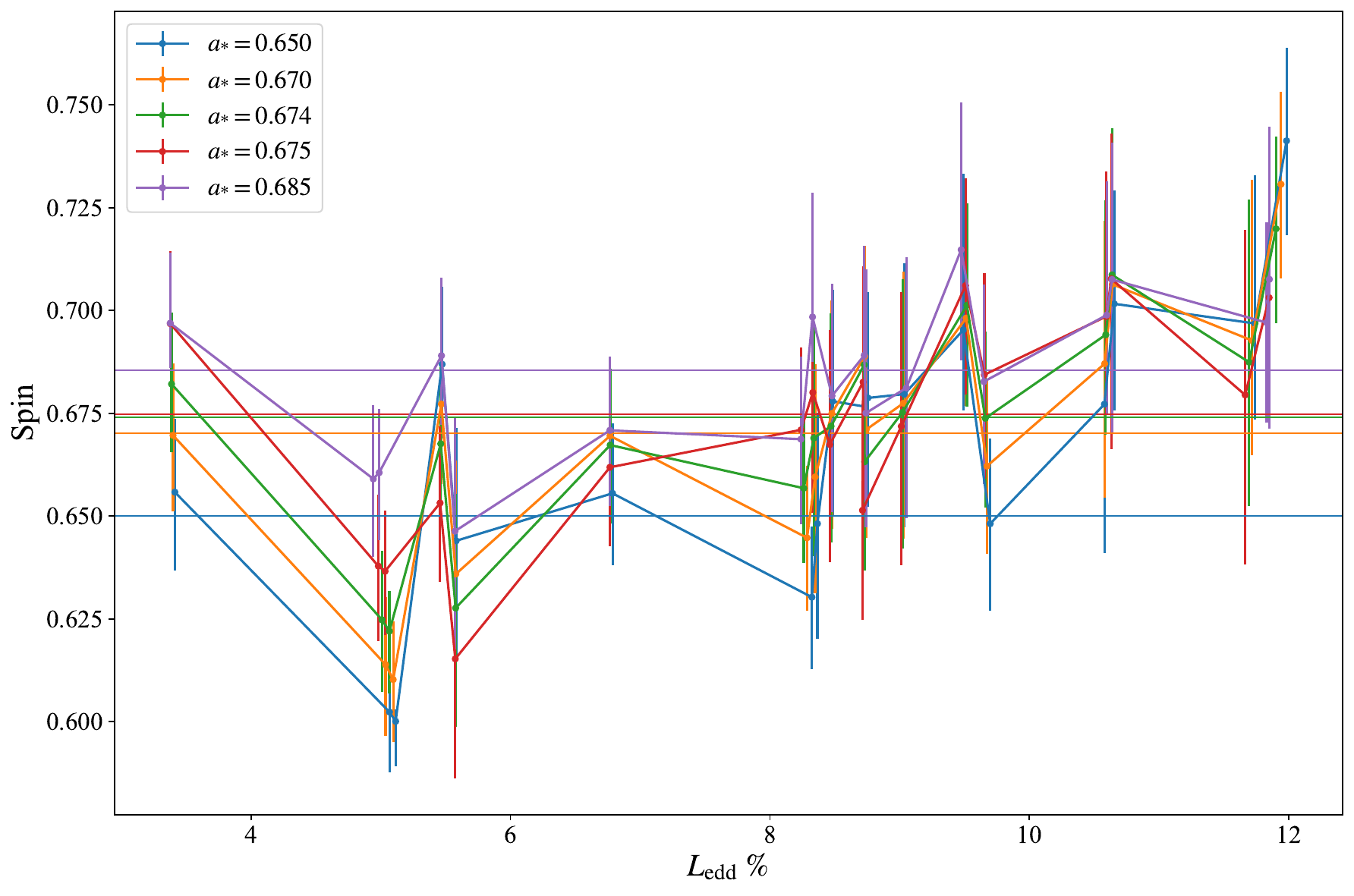}
    \caption{Parameter evolution of \jj from Model 3: {\tt\string tbvarabs(simplcut*kerrbb2)} at different $\Gamma$ values: 2.0, 2.2, 2.4, 2.7 and free $\Gamma$, under fiducial conditions:  $M=10\,M_{\odot}$, $i=40^{\circ}$ and $D=10$\,kpc. Solid lines connecting the points indicate the source's time evolution under different $\Gamma$ values. We have labeled the spin averages for each case (vertical line).}
    \label{fig:spin}
\end{figure}

Given that $\Gamma$ is only weakly constrained in the soft state, we explored the effect of fixing its value in our analysis. We tested its influence on the fitting results by fixing $\Gamma$ at different values. We set $\Gamma$ to 2.0, 2.2, 2.4, and 2.7, respectively, and also performed a fit with $\Gamma$ as a free parameter under a fiducial case $M=10\,M_{\odot}$, $i=40^{\circ}$ and $D=10$\,kpc. Fig.~\ref{fig:spin} shows the spin variations for Model 3 at these values. The corresponding spin are 0.650, 0.670, 0.674, 0.675, and 0.685 (for free $\Gamma$), indicating a small variation of $\Delta a_{*}=0.035$ across all cases. The weighted average of the spin parameter obtained from fits with different values of gamma is $0.669\pm0.035$ and a standard deviation of $\sigma_{a*}=0.035$. This systematic difference is small compared to the expected uncertainty associated with the system parameters $M$, $i$, and $D$, even under the optimistic assumption that these parameters are known to better than $\sim$10\% \citep{gou2011extreme}. This indicates that our spin measurement is relatively insensitive to the exact choice of $\Gamma$, and is therefore robust against reasonable uncertainties in the spectral slope. 

\subsection{The influence of the effect of self-irradiation and limb-darkening}

We also considered the impact of self-irradiation and limb-darkening effects in {\tt\string kerrbb2} on the fitting results. This time, we set both flags to 0 to turn off the effect of self-irradiation and limb-darkening, and obtained a spin value $a_{*}=0.649\pm0.004$ and a standard deviation of $\sigma_{a*}=0.020$. Taking both effects into account, the resulting average spin parameter is $0.667\pm0.026$ with a standard deviation of $\sigma_{a*}=0.026$. Compared with the results that include these two effects, $\Delta a_{*}=0.032$, this variation is small relative to the range typically expected from $\sim$10\% uncertainties in system parameters, and we believe that these effects do not significantly impact the spin value. However, including these two effects should better reflect the actual physical processes.

\subsection{The influence of viscosity parameter \texorpdfstring{$\alpha$}{alpha}}

\begin{figure}[htbp]
    \centering
    \includegraphics[width=0.8\textwidth]{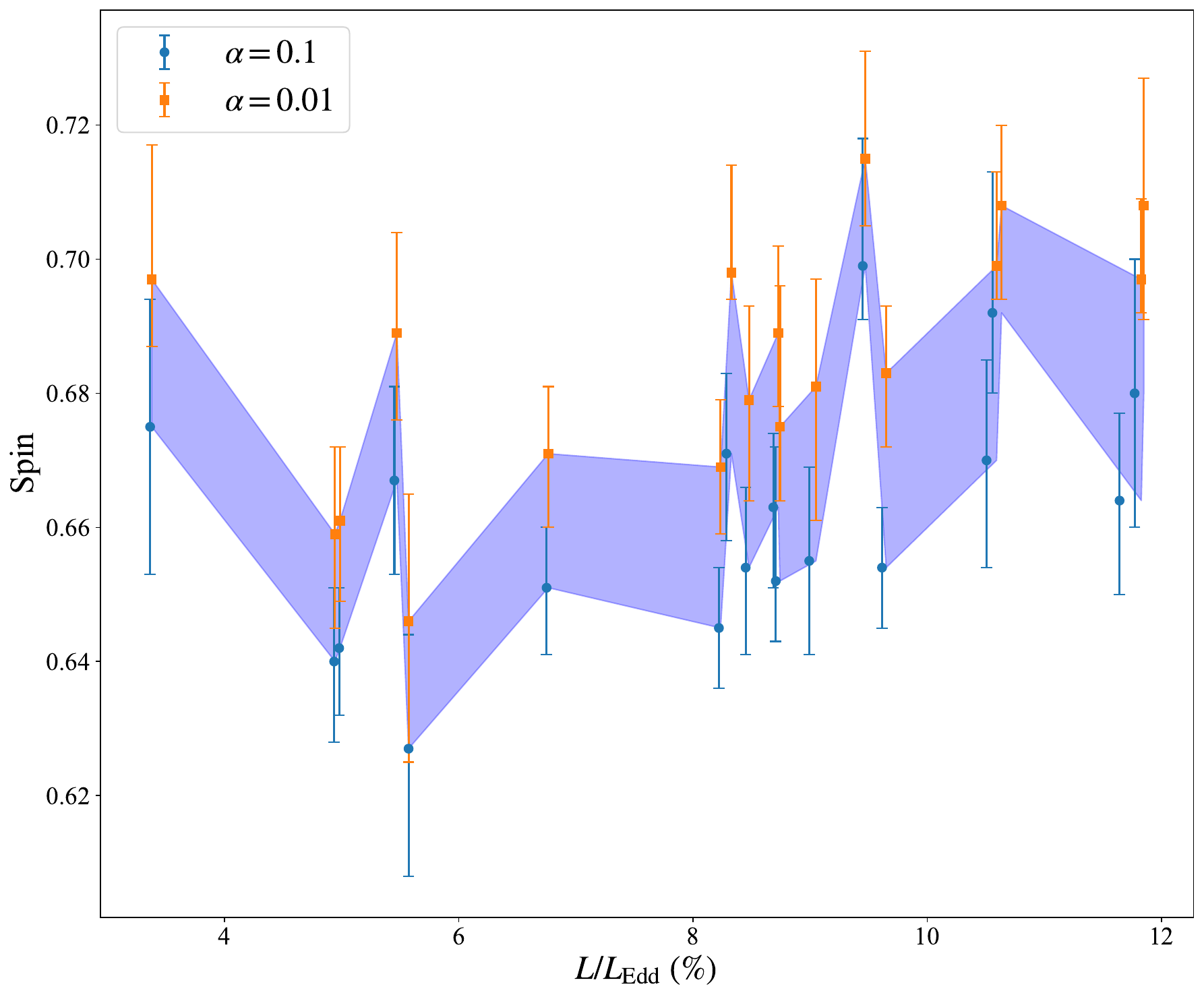}
    \caption{Difference between different $\alpha$. Fitting results with Model 3: {\tt\string tbvarabs(simplcut*kerrbb2)} for different viscosity parameters $\alpha$. Orange points represent $\alpha=0.1$, and blue points represent $\alpha=0.01$. The shaded region indicates the impact on the spin values from assuming either of the two viscosity values.  Note the narrow  range of spin plotted here.  This illustrate that viscosity has a small but systematic impact on the spin inference.}
    \label{fig:alpha}
\end{figure}

In this work, we use the results with $\alpha=0.01$ as the default, while also performing the same calculations with $\alpha=0.1$. Compared to $\alpha=0.01$, we found that $\alpha=0.1$ results in a lower spin (for $\alpha=0.01$, the weighted average spin is $0.685\pm0.004$  and a standard deviation of $\sigma_{a*}=0.017$; for $\alpha=0.1$, the weighted average spin is $0.657\pm0.004$   and a standard deviation of $\sigma_{a*}=0.015$). The variation for these two cases is $\Delta a_{*}=0.028$, which is small compared to the uncertainty in spin resulting from nominal $\sim$10\% uncertainties in key input parameters such as $M$, $i$, and $D$. Table~\ref{tab:alpha0.1} summarizes Model 3 fits assuming $\alpha=0.1$, while Fig.~\ref{fig:alpha} demonstrates the viscosity-dependent spin variation (comparing $\alpha$=0.1 and 0.01).
The dimensionless viscosity parameter ($\alpha$) systematically influences the spin estimate through a well-defined chain of dependencies: a higher $\alpha$ reduces the disk density at a given accretion rate, which increases the color correction factor, leading to a larger inferred inner disk radius and consequently, a lower inferred black hole spin. This variation is consistent with expectations based on analyses of other systems \citep{gou2009determination,feng2023using}. We adopt an averaging over $\alpha=0.01$ and $\alpha=0.1$ for the final result. The weighted average spin, including the intrinsic scatter between the two measurements, is $0.671\pm0.022$, with a corresponding standard deviation of $\sigma_{a*}=0.021$. This is in-line with the findings of \citet{mondal2024first} with the {\tt\string kerrbb} model with {\em IXPE} and {\em NuSTAR} data for similar parameters. However, our spin measurement is lower than that reported by \citet{pengInsightHXMTNICERNuSTAR2024}. Interestingly, the reflection spectroscopy method appears to yield systematically higher spin estimates in several black hole systems \citep{Draghis2025}, such as 4U 1543$-$47 \citep{shafee2005estimating, draghis2023systematic}, GRO J1655$-$40 \citep{shafee2005estimating, reis2009determining}, and GX 339$-$4 \citep{reis2008systematic, kolehmainen2010limits}. This trend is supported by recent larger-sample studies, which indicate that reflection-based methods generally return higher spin values compared to continuum-fitting approaches, suggesting a potential systematic bias that merits further investigation.

\subsection{The influence of the systematic parameters}

Previous studies of \jj have yielded differing constraints on its system parameters. For example, \citet{pengInsightHXMTNICERNuSTAR2024} analyzed  hard state spectra of \jj with {\em Insight-HXMT}, {\em NuSTAR} and {\em NICER} data.  They obtained a low inclination estimate of $21.1^{+4.5}_{-3.6}$ degrees, and adopted a relationship between the total radiated energy and the peak luminosity to estimate a distance of $5.8\pm2.5$~kpc and derived masses of $3.67\pm1.79\,M_{\odot}$ to $8.07\pm4.20\,M_\odot$ with a corresponding spin of $0.84^{+0.17}_{-0.26}$. However, \citet{mondal2024first} provide constraints on the mass of the central black hole, ranging from approximately $9.2\pm1.6$ to $10.5\pm1.8\,M_{\odot}$, with a moderate spin parameter of around $0.6\pm0.2$ to $0.7\pm0.2$ by assuming a distance at 10~kpc. It is worth noting that the results presented here are derived from spectral fitting and do not take the polarization model into account \citep{mondal2024first}. The disk inclination is found to be between $38^{\circ}\pm9^{\circ}$ and $47^{\circ}\pm15^{\circ}$. These discrepancies highlight the need for a systematic exploration. To resolve these discrepancies, our study systematically investigates how spin measurements depend on parameter variations across allowed space, particularly focusing on degeneracy breaking through joint $\chi^{2}$-Eddington ratio constraints.

\begin{figure}[htbp]
    \centering
    \includegraphics[width=\textwidth]{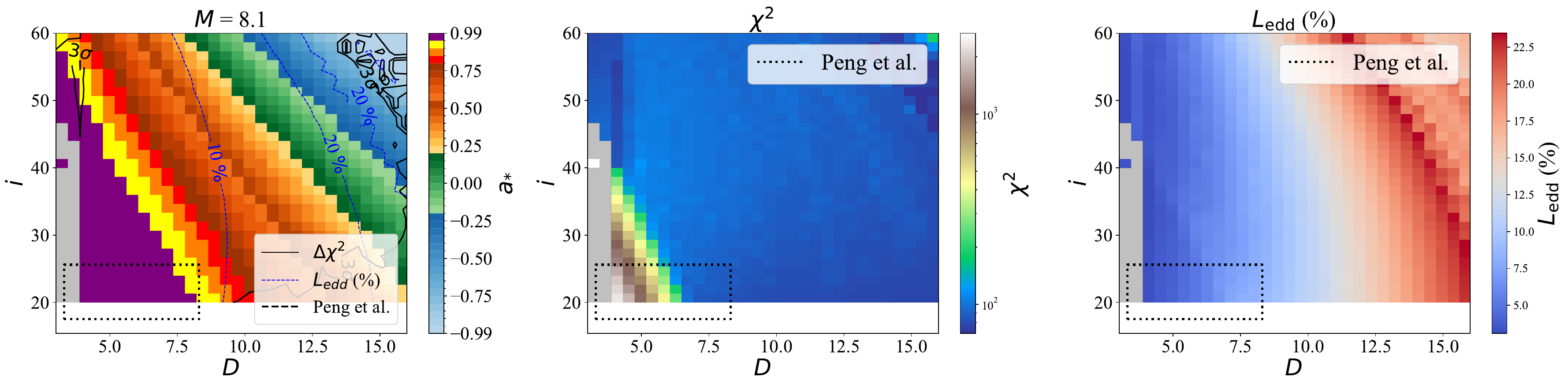}
    \label{a}
    \includegraphics[width=\textwidth]{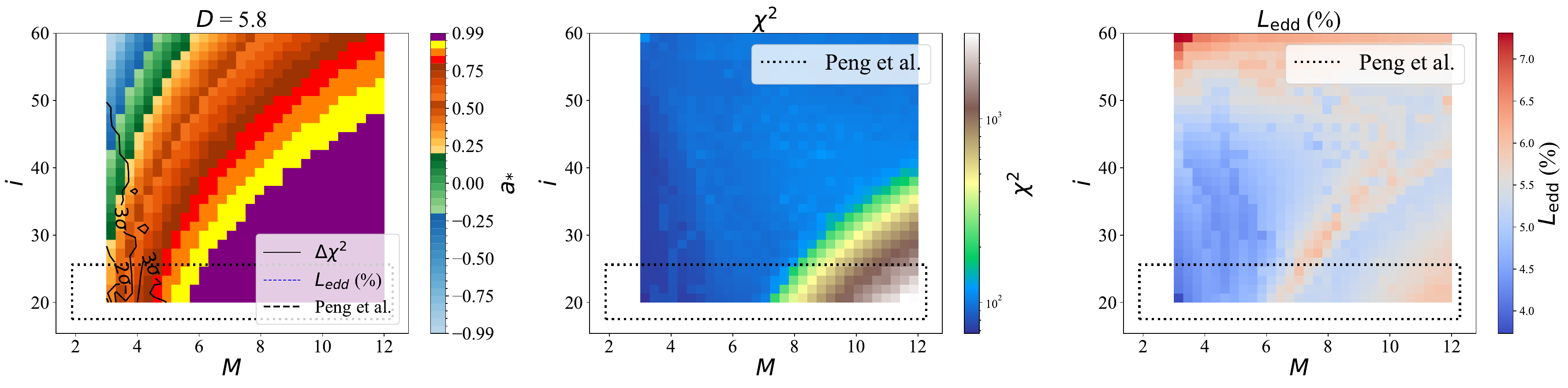}
    \label{b}
    \includegraphics[width=\textwidth]{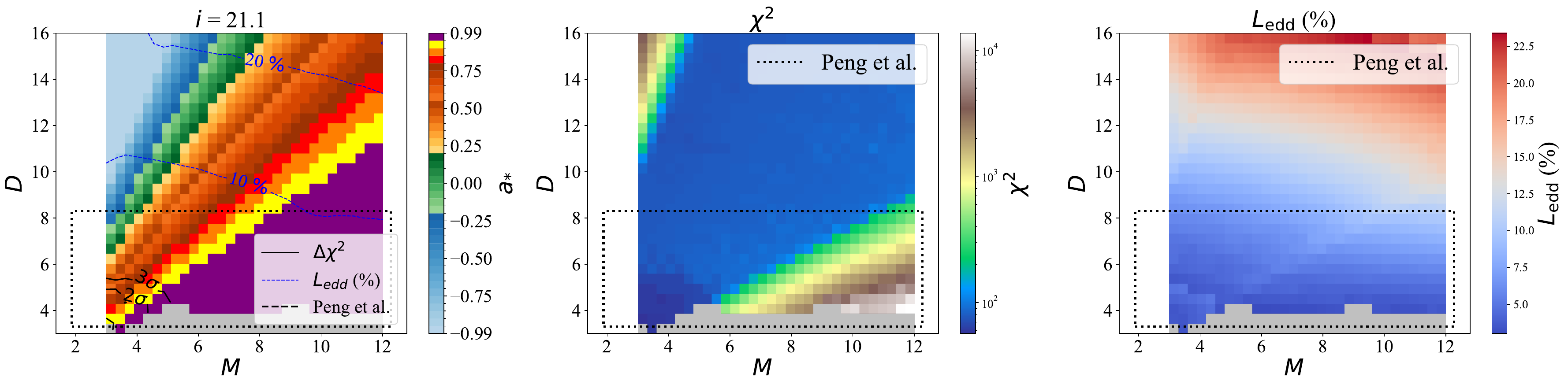}
    \label{c}
    \caption{Parameter space analysis of the \jj. The fitting results are shown with Model 3: {\tt\string tbvarabs*(simplcut*kerrbb2)}. Left: Joint distributions of delta chi-squared ($\Delta\chi^{2}$) (black solid contours at $1\sigma$, $2\sigma$, $3\sigma$) and Eddington-scaled luminosity (blue dashed contours at 10\%, 20\% Eddington). For better visualization, $\Delta\chi^2$ is computed with respect to the minimum chi-squared value. Middle: Two-dimensional projection of the total chi-squared landscape across the parameter space, with dark blue regions indicating better fits. Right: Dependence of Eddington ratio $L/L_{\rm Edd}$ on system parameters, color-coded by the median value at each grid point. The gray-shaded regions mark excluded parameter space where $\chi^{2}/\nu>400$ (see Sect.~\ref{subsec:error} for selection criteria). The parameter space obtained by \citet{pengInsightHXMTNICERNuSTAR2024} is indicated by a dotted line.}
    \label{fig:grid_all}
\end{figure}

Since no {\em dynamical} measurements of mass, inclination, or distance are yet available, we adopted plausible ranges informed by recent studies and existing constraints from Galactic black hole X-ray binary (BHXB) populations: $M=3-12\,M_{\odot}$, $i=20^{\circ}-60^{\circ}$, and $D=3-16$ kpc. We constructed a $30\times30\times30$ grid in this parameter space, applying Model 3, {\tt\string tbvarabs*(simplcut*kerrbb)}, and focus on the orbital night spectra to avoid complications due the NICER light leak from 2023 May 22 onward\footnote{\url{https://heasarc.gsfc.nasa.gov/docs/nicer/analysis_threads/light-leak-overview/}}. Only spectra with a disk fraction $>0.8$ and scattering fraction $f_{\rm sc}<0.25$ were used, ensuring a thermally dominated state \citep{steiner2009simple}. This process yielded 627,102 successful fits, of which 470,063 satisfied the additional filter $\chi^{2}/\nu<400$, minimizing fit-quality artifacts. Some of our data achieve fit quality with $\chi^{2}_{\nu}$ as low as 0.4 which is indicative of having overestimated the systematic errors attributed to detector calibration, e.g., a choice of 1\% instead of the NICERDAS recommended and default value of 1.5\% (which we have adopted here).  Aside from rescaling the $\chi^{2}_{\nu}$, such overfitting has insignificant impact on the best-fitting parameter values and does not affect our conclusions, and so we leave it unaltered from its recommended settings. The resulting spin distribution across the 3D grid is visualized in Fig.~\ref{fig:grid_xkbb}. Similar to Fig.~\ref{fig:grid_all}, grid cells without color indicate regions with no qualifying spectra.
We observe a clear trend in the variation of the spin with respect to the parameters: as the inclination decreases, the mass increases and as the distance decreases, the spin tends to increase.

To investigate trends, we plotted 2D slices of the 3D grid at fixed values of distance ($D=10$\,kpc), mass ($M=10\,M_{\odot}$), or inclination ($i=40^{\circ}$), shown in Fig.~\ref{fig:grid_all}. At each grid point, we used the median spin value when multiple spectra matched the selection. The plots in the left column of Fig.~\ref{fig:grid_all} follow a clear physical logic, echoing Sect.~\ref{subsec:ezdiskbb}: higher mass, lower inclination, and smaller distance tend to raise the inferred spin. These effects are primarily geometric, reflecting how the inner radius (and thus ISCO) is constrained by flux and temperature under varying assumptions. The parameters are highly degenerate, and even modest changes can propagate into spin estimates.
The delta chi-squared ($\Delta\chi^{2}$) contours overlaid in this plot indicate preferred regions in parameter space. While these contours suggest nominal best values, we emphasize that such values should not be interpreted as definitive.  For example, when fixing $M=10\,M_{\odot}$ (top row), low distances ($<4$\,kpc) are disfavored. Similarly, for fixed $D=10$\,kpc (middle row), high inclinations ($>45^{\circ}$) yield poor fits. Fixing $i=40^{\circ}$, the contours exclude nearby, large-mass, low-distance combinations.  

\begin{figure}[htbp]
    \centering
    \includegraphics[width=\textwidth]{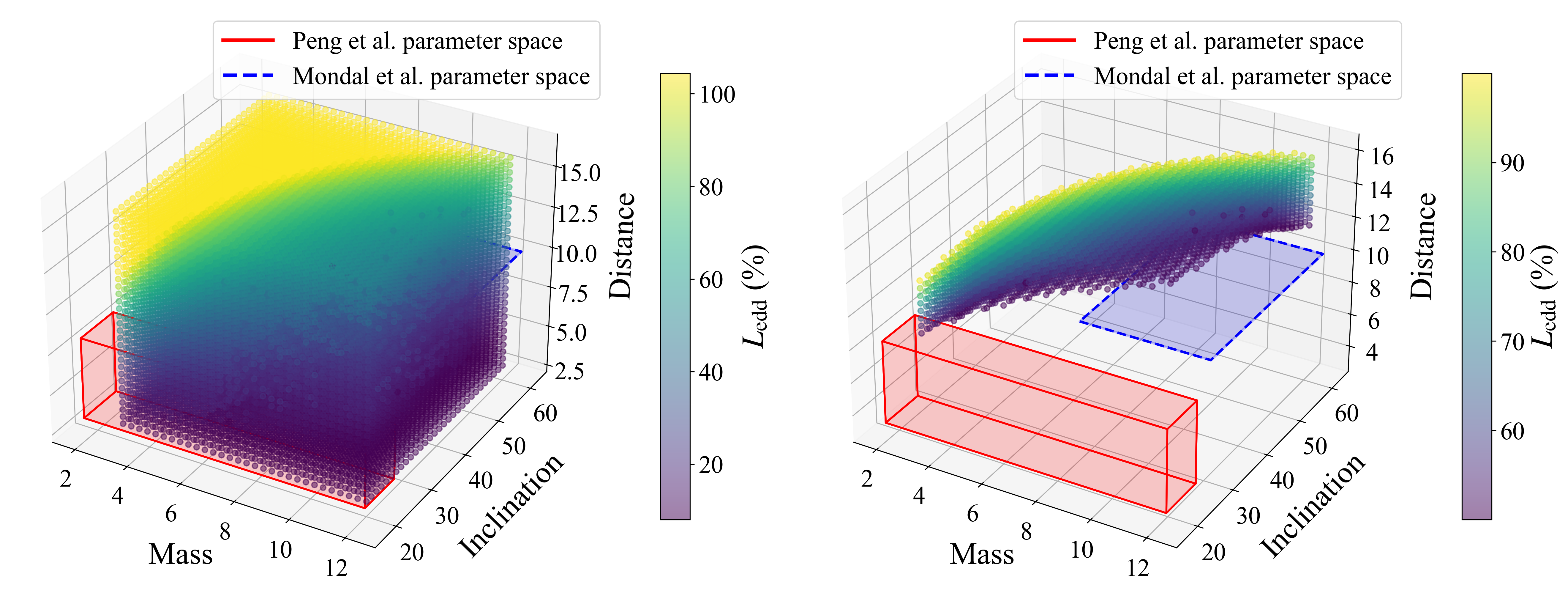}
    \caption{Left panel shows the variation of the peak luminosity (N4) as a function of system parameters: mass, inclination, and distance. To enhance visualization clarity and avoid color saturation, the color map range was limited to the 10th–90th percentile of the data distribution. This excludes extreme values and ensures distinguishable color gradients in the main data range. Right panel presents the parameter space where the peak luminosity lies within the 50\%–100\% range of the Eddington luminosity.}
    \label{fig:peak}
\end{figure}

\begin{figure}[htbp]
    \centering
    \includegraphics[width=\textwidth]{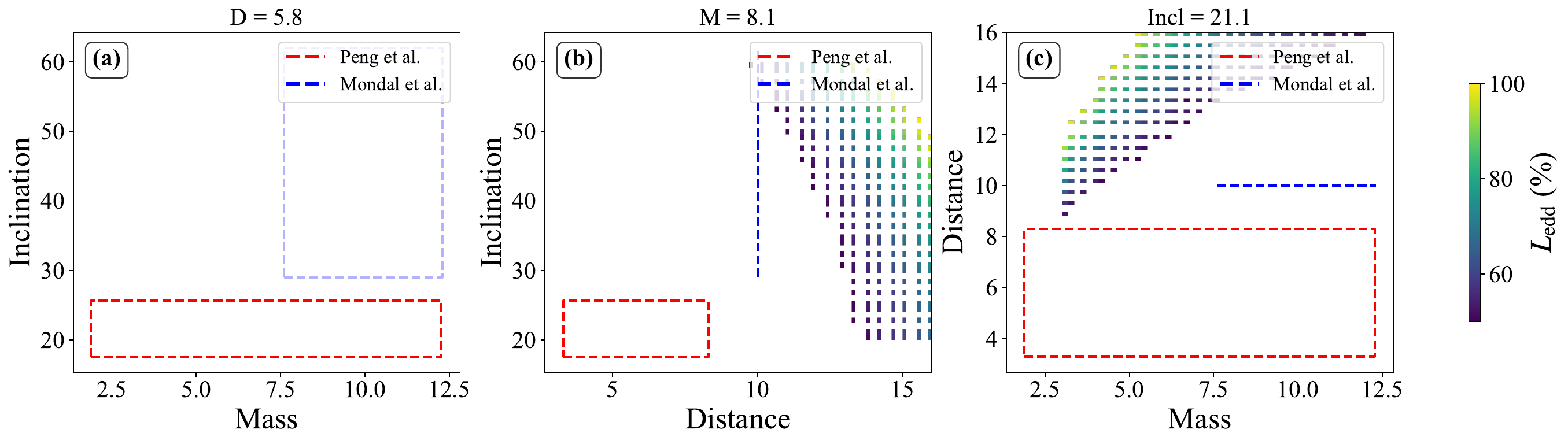}
    \caption{We show two-dimensional slices of the parameter space from \citet{pengInsightHXMTNICERNuSTAR2024}, with the red bounding box marking their constrained parameter ranges. As the distance is fixed at 10 kpc in \citealp{mondal2024first}, the light-blue dashed line in panel (a) only indicates the range of mass and inclination angle. Only results corresponding to peak luminosities between 50\% and 100\% of the Eddington luminosity are displayed.}
    \label{fig:peak_space}
\end{figure}

Further diagnostics of parameter constraints are presented through total $\chi^2$ and Eddington-scaled luminosity (plots in medium and right column of Fig.~\ref{fig:grid_all}). Regions with extremely low masses combined with high inclinations or extremely small distances yield statistically poor fits (high $\chi^2$), supporting constraints derived from other recent studies (e.g., \citealp{mondal2024first, burridge2024hi, pengInsightHXMTNICERNuSTAR2024}). In addition, we plot the variation of the peak luminosity (corresponding to the spectrum N4) across the parameter space. Fig.~\ref{fig:peak} further highlights data points with peak luminosities in the 50\%–100\% range of the Eddington ratio, which is typical for the peak luminosities of X-ray black hole binaries \citep{fender2004towards, tetarenko2016watchdog}. This selection thus provides useful constraints on the system parameters. This conclusion is in tension with the lower-mass and lower-inclination estimates reported by \citet{pengInsightHXMTNICERNuSTAR2024} and \citet{mondal2024first}, suggesting potential systematic differences arising from distinct modeling frameworks (see Fig.~\ref{fig:peak_space}). It is noteworthy, however, that this tension arises specifically within our empirically motivated peak luminosity range; consideration of a broader range (e.g., extending down to $\sim$25\% $L_{\rm Edd}$) would, in fact, reveal regions of overlap with the mass space found by \citet{pengInsightHXMTNICERNuSTAR2024}. We note that the spin and inclination estimates in \citet{pengInsightHXMTNICERNuSTAR2024} were obtained using reflection spectroscopy, in contrast to the continuum-fitting method. It is noteworthy that, unlike the typically high spins derived from reflection spectroscopy in many other black hole X-ray binaries (BHXRBs), the spin values for this particular source obtained via continuum-fitting vs. reflection, do not appear to be in conflict, although a firmer assessment requires precise dynamical measurements. This relative agreement presents a valuable opportunity. A systematic comparison between a sample of BHXRBs like this one, which show consistency between the two methods, and another sample that exhibits significant discrepancies could help isolate the key factors—such as coronal geometry, or data selection criteria—driving the systematic differences. We note that our adopted range of peak luminosity is an empirically motivated expectation, but may not be applicable to this source. Therefore, the present results should be viewed as indicative rather than definitive. Meanwhile, the favored ranges for the black hole mass and inclination angle show clear distance-dependent variations.

In conclusion, while our exploration of the parameter space reveals clear dependencies of spin on the key system parameters, its precise determination remains contingent upon future dynamical measurements of mass, distance, and inclination. The detailed mappings presented in this work—illustrating how spin varies with $M$, $i$ and $D$—establish a versatile and forward-looking framework. Once more accurate constraints on the dynamical parameters become available, our results will allow the community to readily "look up" the corresponding spin value. Furthermore, this approach enables a robust assessment of how different theoretical assumptions or candidate values would influence the spin estimate. We thus anticipate that this systematic and agnostic methodology will prove invaluable for interpreting future observations and for reconciling spin measurements across different techniques.

\section{CONCLUSIONS} \label{sec5}
We conducted a spectral analysis of the black hole transient source \swiftj\ using {\em NICER} data spanning its outburst. This work investigated the source’s evolution during its outburst and employed continuum-fitting to model its spectra in the thermal dominant state in order to constrain the spin. Using  the {\tt\string kerrbb2} model, and adopting a fiducial BH mass of 10~$M_{\odot}$, a distance of 10~kpc, and an inclination angle of 40~degrees, we obtain a moderate spin value of $a_{*}=0.669\pm0.021$. This estimate accounts for all systematic effects discussed in this work but assumes perfect knowledge of the system parameters (i.e., their uncertainties are neglected). However a proper determination of the spin requires better constraints on the system parameters. When accounting for variations of luminosity in the $\alpha=0.01$, the resulting spin value and its standard deviation are: $a_{*}=0.685\pm0.004\,,\sigma_{a*}=0.017$. We present the the spin's dependence over a broader range of possible values for those system parameters which is useful to understand the parameter dependencies in the continuum fitting approach. A future precise dynamical measurement will thereby allow us to make use of the results presented here to obtain a spin measurement from these continuum-fitting data.

\section{ACKNOWLEDGMENTS}

We gratefully acknowledge the Center for Astrophysics $\vert$  Harvard \& Smithsonian for providing essential resources and support. J.F.S. acknowledges support from NASA's NICER General Observer grant 80NSSC25K7130. Y.Z.\ acknowledges support from the Dutch Research Council (NWO) Rubicon Fellowship, file no.\ 019.231EN.021. Y.S. acknowledges the support from the University of Chinese Academy of Sciences Joint Ph.D. Training Program. We thank the anonymous referees for their insightful comments and suggestions, which significantly improved the quality of this manuscript. We also thank the editor for the careful handling of our paper. In addition, we thank Peng and collaborators for fruitful discussions regarding their related work. Y.S. also thanks Yuhao Chen for  useful discussions. This work was supported by the National Natural Science Foundation of China (NSFC) (12273058), and by the National Program on Key Research and Development Project (Grant No. 2023YFA1607900). We acknowledge the science research grants from the China Manned Space Project. NICER launched on June 3, 2017, and is a 0.2--12 keV X-ray telescope on the International Space Station (ISS). This work utilized data and software provided by HEASARC, operated by the Astrophysics Science Division at NASA's Goddard Space Flight Center.

\appendix




\setcounter{table}{0}
\renewcommand{\thetable}{A\arabic{table}}
\section{The results from different viscosity parameter}
\input{table_kerrbb2_alpha0.1}

\FloatBarrier

\bibliography{1_ref}{}
\bibliographystyle{aasjournal}

\end{document}

%% file: table_obs.tex
\rowcolors{42}{gray!20}{gray!20}
\begin{longtable}{|c|c|c|c|c|c|c|}
    \caption{Swift J151857.0$-$572147 \textit{NICER}/XTI observation overview.}
    \label{tab:obs}\\

    \toprule
        \multirow{2}{*}{Number} & \multirow{2}{*}{ObsID} & \multirow{2}{*}{Seg} & \multirow{2}{*}{MJD} & \multirow{2}{*}{Start time} & Exposure & Count Rates \\
        & & & (day) & & (s) & ($\rm{cts\ s^{-1}\ 52\ FPM^{-1}}$) \\
    \midrule
    \endfirsthead
        
    \multicolumn{7}{r}{Continued}\\
    \toprule
        \multirow{2}{*}{Number} & \multirow{2}{*}{ObsID} & \multirow{2}{*}{Seg} & \multirow{2}{*}{MJD} & \multirow{2}{*}{Start time} & Exposure & Count Rates \\
        & & & (day)& & (s) & ($\rm{cts\ s^{-1}\ 52\ FPM^{-1}}$) \\
        \midrule
    \endhead
        
    \bottomrule
    \multicolumn{7}{c}{Continued on next page.}\\
    \endfoot

    \bottomrule
    \endlastfoot
    
    N1 & 7204220101$^{d}$ & w & 60373.24 & 2024-03-04 05:42:06& 2920.48 & 1534.21 \\
    N2 & 6756010102$^{d}$ & w & 60376.14 & 2024-03-07 03:15:11& 1628.08 & 2072.16\\
    N3 & 6756010103$^{d}$ & w & 60377.10 & 2024-03-08 02:24:59& 195.00 & 2239.01 \\
    N4 & 7204220104$^{d}$ & w & 60377.55 & 2024-03-08 13:12:06& 4912.00 & 2146.86 \\
    N5 & 7204220105$^{d}$ & w & 60378.19 & 2024-03-09 04:40:15& 3668.00 & 2176.18 \\
    N6 & 7204220106$^{d}$ & w & 60379.23 & 2024-03-10 05:28:32& 3271.99 & 1936.62 \\
    N7 & 7204220107$^{d}$ & w & 60380.00 & 2024-03-11 00:05:32& 7274.96 & 1775.27\\
    N8 & 7204220108$^{d}$ & w & 60381.10 & 2024-03-12 02:28:29& 4571.00 & 1852.02 \\
    N9 & 7204220109$^{d}$ & w & 60382.01 & 2024-03-13 00:12:55& 3714.00 & 1765.08 \\
    N10 & 7204220110$^{d}$ & w & 60383.04 & 2024-03-14 01:03:01& 1644.98 & 1626.03\\
    N11 & 7204220111 & w & 60387.55 & 2024-03-18 13:09:18& 4196.00 & 1468.45 \\
    N12 & 6756010104 & w & 60389.67 & 2024-03-20 16:02:47& 1266.00 & 1246.10 \\
    N13 & 6756010105$^{d}$ & w & 60390.27 & 2024-03-21 06:25:19& 2526.00 & 1209.41 \\
    N14 & 6756010106$^{d}$ & w & 60391.04 & 2024-03-22 00:57:18& 384.00 & 1285.43 \\
    N15 & 6756010106 & w & 60391.35 & 2024-03-22 08:17:05& 5185.24 & 1214.86 \\
    N16 & 6756010107 & w & 60391.99 & 2024-03-22 23:46:08& 6111.47 & 1161.81 \\
    N17 & 6756010108 & w & 60398.12 & 2024-03-29 02:52:12& 811.00 & 793.88 \\
    N18 & 6756010109 & w & 60400.57 & 2024-03-31 13:42:48& 2213.00 & 808.68 \\
    N19 & 6756010110 & w & 60401.09 & 2024-04-01 02:05:48& 2933.00 & 780.08 \\
    N20 & 6756010111 & w & 60402.06 & 2024-04-02 01:19:29& 2171.00 & 732.13 \\
    N21 & 7661010101 & w & 60404.18 & 2024-04-04 04:25:10& 3907.00 & 669.22 \\
    N22 & 7204220112 & w & 60405.22 & 2024-04-05 05:13:27& 569.00 & 616.01 \\
    N23 & 7204220113 & w & 60408.90 & 2024-04-08 21:40:12& 228.00 & 538.47 \\
    N24 & 7204220114 & w & 60409.16 & 2024-04-09 03:48:48& 404.00 & 528.17 \\
    N25 & 7661010103 & w & 60412.06 & 2024-04-12 01:23:11& 2682.00 & 466.46\\
    N26 & 7204220115 & w & 60412.77 & 2024-04-12 18:27:42& 268.00 & 460.98 \\
    N27 & 7204220116 & w & 60413.67 & 2024-04-13 16:06:27& 779.00 & 430.97 \\
    N28 & 7204220117 & w & 60414.06 & 2024-04-14 01:24:13& 1411.00 & 420.06 \\
    N29 & 7204220118$^{d}$ & w & 60415.41 & 2024-04-15 09:54:06& 85.23 & 407.68 \\
    N30 & 7204220118 & w & 60415.41 & 2024-04-15 09:55:18& 1065.00 & 408.75 \\
    N31 & 7204220119$^{d}$ & w & 60416.25 & 2024-04-16 06:01:33& 90.00 & 406.17 \\
    N32 & 7204220119 & w & 60416.25 & 2024-04-16 06:03:08& 1265.00 & 396.84 \\
    N33 & 7204220120$^{d}$ & w & 60417.28 & 2024-04-17 06:47:08& 301.00 & 378.13\\
    N34 & 7204220120 & 1 & 60417.28 & 2024-04-17 06:49:43& 564.00 & 382.11 \\
    N35 & 7661010104$^{d}$ & w & 60417.41 & 2024-04-17 09:52:47& 471.00 & 390.08 \\ 
    \rowcolors{0}{}{}
    N36 & 7661010104 & 1,3,6,9,10 & 60417.41 & 2024-04-17 09:55:30& 3107.00 & 375.10 \\  
    N37 & 7661010105$^{d}$ & w & 60425.09 & 2024-04-25 02:03:47& 755.00 & 278.63 \\ 
    \rowcolor{gray!20}{}
    N38 & 7661010105 & w & 60425.09 & 2024-04-25 02:06:53& 2476.00 & 286.37 \\
    \rowcolor{gray!20}{}
    N39 & 7661010106 & w & 60432.90 & 2024-05-02 21:30:40& 459.00 & 277.42 \\
    \rowcolor{gray!20}{}
    N40 & 7661010107 & w & 60433.02 & 2024-05-03 00:30:09& 2021.00 & 221.49 \\
    \rowcolor{gray!20}{}
    N41 & 7661010108$^{d}$ & 1 & 60454.14 & 2024-05-24 03:23:55& 66.00 & 180.63\\
    \rowcolor{gray!20}{}
    N42 & 7661010108 & w & 60454.96 & 2024-05-24 22:56:02& 1415.00 & 180.62 \\
    \rowcolor{gray!20}{}
    N43 & 7661010109 & 1,2 & 60455.04 & 2024-05-25 00:53:28& 832.00 & 179.58 \\
    \rowcolor{gray!20}{}
    N44 & 7661010109$^{d}$ & 1 & 60455.04 & 2024-05-25 01:02:10 & 349.24 & 186.83 \\
    \rowcolor{gray!20}{}
    N45 & 7661010110 & w & 60468.45 & 2024-06-07 10:42:57& 785.00 & 102.55 \\
    \rowcolor{gray!20}{}
    N46 & 7661010111$^{d}$ & w & 60475.01 & 2024-06-14 00:15:35& 3224.00 & 80.37 \\
    \rowcolor{gray!20}{}
    N47 & 7661010112$^{d}$ & w & 60482.69 & 2024-06-21 16:37:30& 300.00 & 68.94 \\
    N48 & 7661010112 & w & 60482.69 & 2024-06-21 16:41:05& 75.00 & 72.13 \\
    N49 & 7661010114$^{d}$ & w & 60488.56 & 2024-06-27 13:30:45& 140.00 & 43.58 \\
    N50 & 7661010115$^{d}$ & 1,2,5,6 & 60489.14 & 2024-06-28 03:27:22& 187.00 & 64.57 \\
    N51 & 7661010116$^{d}$ & w & 60496.64 & 2024-07-05 15:15:46& 581.00 & 31.89 \\
    N52 & 7661010117$^{d}$ & w & 60497.02 & 2024-07-06 00:34:51& 903.00 & 31.43 \\
    N53 & 7661010117 & w & 60497.03 & 2024-07-06 00:40:17& 2010.00 & 30.74 \\
    N54 & 7661010118$^{d}$ & w & 60499.46 & 2024-07-08 11:05:27& 1897.00 & 31.00 \\
    N55 & 7661010120$^{d}$ & w & 60503.01 & 2024-07-12 00:15:54& 1149.00 & 27.53 \\
    N56 & 7661010121$^{d}$ & w & 60507.98 & 2024-07-16 23:31:05& 456.00 & 18.74 \\
    N57 & 7661010122$^{d}$ & w & 60508.04 & 2024-07-17 01:04:02& 2004.98 & 18.34 \\
    N58 & 7661010123 & 1,3,8 & 60518.02 & 2024-07-27 00:26:59& 619.00 & 16.14 \\
    N59 & 7661010124$^{d}$ & w & 60531.36 & 2024-08-09 08:41:02 & 1441.54 & 8.58 \\      
    N60 & 7661010124 & w & 60531.37 & 2024-08-09 08:49:25& 2713.00 & 8.97 \\
    N61 & 7661010127 & w & 60544.53 & 2024-08-22 12:48:47& 2199.00 & 7.88 \\

\end{longtable}
\parbox{\textwidth}{\footnotesize \textit{Notes}: summarizes the {\em NICER/XTI}  observations of Swift J151857.0$-$572147. Columns sequentially indicate: (1) spectrum identification number; (2) Observation ID (ObsID). The superscript ``d'' in the upper right corner indicates that the data corresponds to orbit day, the remainder consists of orbit-night data; (3) the index number of GTI segments retained (i.e., only specified in cases where some high-background time-intervals were excluded), ``w'' indicating whole observations with no GTI screening; (4) Modified Julian Date (MJD); (5) observation start time; (6) exposure duration in seconds (s); (7) total count rate in the 0.5-10 keV energy range, normalized to 52 focal plane modules (FPMs; cts/s/52FPM). Observations optimal for continuum fitting analyses—defined by disk fractions exceeding 0.8 and luminosities within 3–30\% of the Eddington limit—are highlighted in gray.}

%% file: table_joint.tex
\begin{table}[ht]
    \centering
    \caption{Joint-fitting of six spectra}
    \label{tab:joint}
    \begin{tabular}{|c|c|ccc|cc|ccc|}
        \toprule
         \multirow{2}{*}{Number} & \multirow{2}{*}{ObsID} & \multicolumn{3}{c|}{$\mathtt{tbvarabs}$} & \multicolumn{2}{c|}{$\mathtt{ezdiskbb}$} & \multicolumn{3}{c|}{$\mathtt{nthcomp}$} \\
        \cmidrule(lr){3-5}  \cmidrule(lr){6-7} \cmidrule(lr){8-10}
        & &$N_{H}\ (10^{22}\ \rm{cm}^{-2})$ & Fe(Mg) & Si & $T_{max}\ (\rm{keV})$ & $N_{disk}$ & $\Gamma$ & $kT_{e}\ (\rm{keV})$ & $N_{nthcomp}$ \\
        \toprule
        N11 & 7204220111  &\multirow{6}{*}{$5.06^{+0.12}_{-0.12}$} & \multirow{6}{*}{$1.81^{+0.16}_{-0.15}$} &  \multirow{6}{*}{$2.29^{+0.11}_{-0.11}$} &  $0.81^{+0.02}_{-0.01}$ &$374.5^{+22.0}_{-23.8}$ & $3.38^{p}_{-0.20}$ & \multirow{6}{*}{100.0$^{f}$} &  $2.46^{+0.44}_{-0.50}$ \\ 
        N22 & 7204220112 &&  &  & $0.70^{+0.01}_{-0.01}$ & $517.9^{+28.9}_{-27.5}$ &  $2.75^{+0.56}_{-0.54}$ &  &  $0.28^{+0.23}_{-0.14}$ \\
        N38 & 7661010105 &  &  &&  $0.60^{+0.01}_{-0.01}$ &$546.4^{+21.2}_{-26.8}$ & $3.50^{p}_{-0.28}$ &  &  $0.18^{+0.01}_{-0.05}$ \\ 
        N45 & 7661010110 &  & &  & $0.50^{+0.01}_{-0.01}$ &$561.9^{+39.2}_{-41.6}$ &  $3.17^{p}_{p}$ &  &  $0.04^{+0.03}_{-0.04}$ \\
        N53 & 7661010117 &  &  &&  $0.409^{+0.004}_{-0.004}$ &$624.4^{+47.8}_{-44.0}$ &  $3.50^{p}_{-0.69}$ &  &  $0.012^{+0.004}_{-0.007}$ \\ 
        N60 & 7661010124  & & &  &  $0.32^{+0.01}_{-0.01}$ &$824.9^{+200.2}_{-146.5}$ & $2.14^{+1.18}_{p}$ &  &  $0.004^{+0.014}_{-0.003}$ \\ 
        \toprule
        $\chi^2/\nu$ & \multicolumn{9}{c|}{416.99/561=0.74}\\
        \bottomrule
    \end{tabular}
    \parbox{\textwidth}{\footnotesize \textit{Notes}: The superscript ``f'' indicates that the parameter is fixed, and ``p'' indicates that the upper and lower limits provided by the fit have reached the boundary constraints. The abundance is relative to Solar.}
\end{table}

%% file: table_xkbb.tex
\rowcolors{0}{}{}
\begin{longtable}{|c|c|cc|ccc|c|c|c|}
    \caption{model 3:\texttt{tbvarabs*(simplcut*kerrbb2)}}
    \label{tab:kerrbb2}\\

    \toprule
        \multirow{2}{*}{Numer} & \multirow{2}{*}{Obs} & \multicolumn{2}{c|}{$\mathtt{simplcut}$} & \multicolumn{3}{c|}{$\mathtt{kerrbb2}$} & \multirow{2}{*}{$\chi^{2}/\nu$} & \multirow{2}{*}{$L/L_{\rm Edd}(\%)$} \\
        \cmidrule(lr){3-4}  \cmidrule(lr){5-7}
        & & $\Gamma$ & $f_{\rm sc}$ & $a_{*}$ & $\dot{M} (10^{18}\;{\rm g}\;{\rm s}^{-1})$ & $f_{\rm{col}}$ & & \\
    \midrule
    \endfirsthead
        
    \multicolumn{8}{r}{Continued}\\
    \toprule
        \multirow{2}{*}{Numer} & \multirow{2}{*}{Obs} & \multicolumn{2}{c|}{$\mathtt{simplcut}$} & \multicolumn{3}{c|}{$\mathtt{kerrbb2}$}  & \multirow{2}{*}{$\chi^{2}/\nu$} & \multirow{2}{*}{$L/L_{\rm Edd}(\%)$} \\
        \cmidrule(lr){3-4}  \cmidrule(lr){5-7}
         & & $\Gamma$ & $f_{\rm sc}$ & $a_{*}$ & $\dot{M} (10^{18}\;{\rm g}\;{\rm s}^{-1})$ & $f_{\rm{col}}$ & &\\
    \midrule
    \endhead

    \bottomrule
    \multicolumn{7}{c}{Continued on next page}\\
    \endfoot

    \bottomrule
    \endlastfoot

N21 & 7661010101 & $2.64^{+0.23}_{-0.26}$ & $0.100^{+0.030}_{-0.027}$ & $0.708^{+0.017}_{-0.019}$ & $1.584^{+0.037}_{-0.036}$ & 1.65 & 0.521 & 11.85 \\
N22 & 7204220112 & $2.00^{+0.18}_{p}$ & $0.019^{+0.005}_{-0.001}$ & $0.697^{+0.005}_{-0.012}$ & $1.593^{+0.024}_{-0.024}$ & 1.65 & 1.093 & 11.83 \\
N23 & 7204220113 & $2.23^{+0.77}_{-0.23}$ & $0.010^{+0.011}_{p}$ & $0.708^{+0.014}_{-0.012}$ & $1.422^{+0.033}_{-0.037}$ & 1.64 & 1.147 & 10.64 \\
N24 & 7204220114 & $2.58^{p}_{-0.22}$ & $0.010^{+0.016}_{p}$ & $0.699^{+0.005}_{-0.014}$ & $1.434^{+0.033}_{-0.024}$ & 1.64 & 0.949 & 10.59 \\
N25 & 7661010103 & $2.59^{+0.17}_{-0.12}$ & $0.010^{+0.002}_{p}$ & $0.683^{+0.011}_{-0.010}$ & $1.336^{+0.024}_{-0.025}$ & 1.64 & 1.414 & 9.65 \\
N26 & 7204220115 & $3.50^{p}_{-0.56}$ & $0.010^{+0.013}_{p}$ & $0.715^{+0.010}_{-0.016}$ & $1.255^{+0.036}_{-0.027}$ & 1.64 & 1.085 & 9.47 \\
N27 & 7204220116 & $2.00^{+0.29}_{p}$ & $0.011^{+0.004}_{p}$ & $0.681^{+0.020}_{-0.016}$ & $1.253^{+0.032}_{-0.032}$ & 1.63 & 0.930 & 9.05 \\
N28 & 7204220117 & $2.14^{+0.42}_{-0.14}$ & $0.026^{+0.018}_{-0.005}$ & $0.675^{+0.011}_{-0.021}$ & $1.223^{+0.035}_{-0.028}$ & 1.63 & 0.747 & 8.74 \\
N29 & 7204220118$^{d}$ & $3.18^{p}_{-0.79}$ & $0.010^{p}_{p}$ & $0.710^{+0.031}_{-0.025}$ & $1.162^{+0.048}_{-0.036}$ & 1.63 & 0.963 & 8.72 \\
N30 & 7204220118 & $2.20^{+0.23}_{-0.10}$ & $0.010^{+0.003}_{p}$ & $0.689^{+0.011}_{-0.013}$ & $1.198^{+0.027}_{-0.023}$ & 1.63 & 1.014 & 8.73 \\
N31 & 7204220119$^{d}$ & $2.11^{+0.82}_{-0.11}$ & $0.010^{p}_{p}$ & $0.705^{+0.012}_{-0.024}$ & $1.143^{+0.048}_{-0.021}$ & 1.63 &  0.952 & 8.49 \\
N32 & 7204220119 & $2.00^{+0.11}_{*}$ & $0.012^{+0.002}_{-0.001}$ & $0.679^{+0.015}_{-0.014}$ & $1.179^{+0.027}_{-0.028}$ & 1.63 & 1.276 & 8.48 \\
N33 & 7204220120$^{d}$ & $2.71^{+0.68}_{-0.41}$ & $0.010^{p}_{p}$ & $0.706^{+0.014}_{-0.012}$ & $1.095^{+0.027}_{-0.033}$ & 1.63 & 1.341 & 8.17 \\
N34 & 7204220120 & $3.50^{p}_{-0.34}$ & $0.010^{+0.003}_{p}$ & $0.698^{+0.004}_{-0.016}$ & $1.128^{+0.030}_{-0.011}$ & 1.63 & 1.334 & 8.33 \\
N35 & 7661010104$^{d}$ & $2.00^{+1.13}_{p}$ & $0.010^{+0.022}_{p}$ & $0.706^{+0.013}_{-0.017}$ & $1.108^{+0.031}_{-0.028}$ & 1.63 & 1.224 & 8.26 \\
N36 & 7661010104 & $2.65^{+0.16}_{-0.12}$ & $0.010^{+0.001}_{p}$ & $0.669^{+0.010}_{-0.010}$ & $1.162^{+0.020}_{-0.021}$ & 1.63 & 1.868 & 8.24 \\
N38 & 7661010105 & $2.22^{+0.27}_{-0.06}$ & $0.010^{+0.004}_{p}$ & $0.671^{+0.011}_{-0.010}$ & $0.952^{+0.018}_{-0.018}$ & 1.62 & 0.897 & 6.76 \\
N39 & 7661010106 & $2.00^{+0.12}_{p}$ & $0.028^{+0.005}_{-0.002}$ & $0.646^{+0.021}_{-0.019}$ & $0.809^{+0.027}_{-0.029}$ & 1.60 & 0.887 & 5.57 \\
N40 & 7661010107 & $2.00^{+0.18}_{p}$ & $0.022^{+0.006}_{-0.001}$ & $0.689^{+0.013}_{-0.015}$ & $0.750^{+0.019}_{-0.017}$ & 1.60 & 0.626 & 5.47 \\
N41 & 7661010108$^{d}$ & $3.50^{p}_{-1.50}$ & $0.010^{+0.057}_{p}$ & $0.663^{+0.046}_{-0.078}$ & $0.719^{+0.078}_{-0.060}$ & 1.59 & 0.831 & 5.06 \\
N42 & 7661010108 & $3.50^{p}_{-0.07}$ & $0.010^{+0.001}_{p}$ & $0.661^{+0.012}_{-0.011}$ & $0.711^{+0.016}_{-0.016}$ & 1.59 & 1.411 & 4.99 \\
N43 & 7661010109 & $3.50^{p}_{-0.16}$ & $0.010^{+0.001}_{p}$ & $0.659^{+0.014}_{-0.013}$ & $0.706^{+0.018}_{-0.019}$ & 1.59 & 1.460 & 4.94 \\
N44 & 7661010109$^{d}$ & $2.24^{p}_{-0.24}$ & $0.010^{+0.027}_{p}$ & $0.673^{+0.029}_{-0.028}$ & $0.680^{+0.034}_{-0.034}$ & 1.59 & 0.857 & 4.85 \\
N45 & 7661010110 & $2.71^{+0.70}_{-0.19}$ & $0.010^{+0.010}_{p}$ & $0.697^{+0.010}_{-0.020}$ & $0.458^{+0.017}_{-0.011}$ & 1.56 & 0.901 & 3.38 \\
N46 & 7661010111$^{d}$ & $3.49^{p}_{-0.26}$ & $0.010^{p}_{p}$ & $0.699^{+0.003}_{-0.010}$ & $0.397^{+0.009}_{-0.006}$ & 1.54 & 1.001 & 2.94 \\
N47 & 7661010112$^{d}$ & $2.48^{p}_{-0.48}$ & $0.010^{p}_{p}$ & $0.709^{+0.033}_{-0.036}$ & $0.325^{+0.044}_{-0.021}$ & 1.51 & 0.865 & 2.44 \\

\end{longtable}
\parbox{\textwidth}{\footnotesize \textit{Notes}: The fitting result with Model 3:  {\tt\string tbvarabs(simplcut*kerrbb2)} and the viscosity parameter $\alpha=0.01$. The superscript ``d'' denotes data from orbit day, and ``p'' indicates that the upper and lower limits provided by the fit have reached the boundary constraints.}

%% file: table_kerrbb2_alpha0.1.tex
\begin{table}[htbp]
\centering
\caption{model 3:\texttt{tbvarabs*(simplcut*kerrbb2)}}
\label{tab:alpha0.1}
\rowcolors{0}{}{}
\begin{tabular}{|c|c|cc|ccc|c|c|}
    \toprule
    \multirow{2}{*}{Number} & \multirow{2}{*}{Obs} & \multicolumn{2}{c|}{$\mathtt{simplcut}$} & \multicolumn{3}{c|}{$\mathtt{kerrbb2}$} & \multirow{2}{*}{$\chi^{2}/\nu$} & \multirow{2}{*}{$L/L_{\rm Edd}(\%)$} \\
    \cmidrule(lr){3-4}  \cmidrule(lr){5-7}
     & & $\Gamma$ & $f_{\rm sc}$ & $a_{*}$ & $\dot{M} (10^{18}\;{\rm g}\;{\rm s}^{-1})$ & $f_{\rm{col}}$ & & \\
    \midrule
            
N21 & 7661010101 & $2.61^{+0.22}_{-0.22}$ & $0.099^{+0.028}_{-0.022}$ & $0.680^{+0.020}_{-0.020}$ & $1.636^{+0.039}_{-0.040}$ & 1.67 & 0.495 & 11.77 \\
N22 & 7204220112 & $2.00^{+0.20}_{p}$ & $0.019^{+0.006}_{-0.002}$ & $0.664^{+0.014}_{-0.013}$ & $1.653^{+0.036}_{-0.037}$ & 1.67 & 1.073 & 11.64 \\
N23 & 7204220113 & $2.31^{+0.70}_{-0.28}$ & $0.010^{+0.008}_{p}$ & $0.692^{+0.012}_{-0.021}$ & $1.441^{+0.047}_{-0.030}$ & 1.67 & 1.103 & 10.56 \\
N24 & 7204220114 & $2.60^{+0.90}_{-0.26}$ & $0.010^{+0.017}_{p}$ & $0.670^{+0.016}_{-0.015}$ & $1.481^{+0.036}_{-0.037}$ & 1.66 & 0.936 & 10.51 \\
N25 & 7661010103 & $2.57^{+0.18}_{-0.12}$ & $0.010^{+0.002}_{p}$ & $0.654^{+0.009}_{-0.009}$ & $1.382^{+0.021}_{-0.022}$ & 1.66 & 1.390 & 9.61 \\
N26 & 7204220115 & $3.35^{+0.15}_{-0.73}$ & $0.010^{+0.007}_{p}$ & $0.699^{+0.008}_{-0.019}$ & $1.278^{+0.038}_{-0.023}$ & 1.66 & 1.039 & 9.45 \\
N27 & 7204220116 & $2.00^{+0.32}_{p}$ & $0.011^{+0.005}_{p}$ & $0.655^{+0.014}_{-0.014}$ & $1.291^{+0.030}_{-0.030}$ & 1.65 & 0.902 & 8.99 \\
N28 & 7204220117 & $2.13^{+0.47}_{-0.13}$ & $0.026^{+0.020}_{-0.005}$ & $0.652^{+0.009}_{-0.020}$ & $1.255^{+0.034}_{-0.029}$ & 1.65 & 0.718 & 8.71 \\
N30 & 7204220118 & $2.19^{+0.26}_{-0.10}$ & $0.010^{+0.004}_{p}$ & $0.663^{+0.012}_{-0.011}$ & $1.235^{+0.025}_{-0.025}$ & 1.65 & 1.009 & 8.69 \\
N32 & 7204220119 & $2.00^{+0.12}_{p}$ & $0.012^{+0.002}_{-0.001}$ & $0.654^{+0.013}_{-0.012}$ & $1.216^{+0.025}_{-0.026}$ & 1.65 & 1.262 & 8.45 \\
N34 & 7204220120 & $3.50^{p}_{-0.34}$ & $0.010^{+0.003}_{p}$ & $0.671^{+0.013}_{-0.012}$ & $1.166^{+0.026}_{-0.027}$ & 1.65 & 1.354 & 8.29 \\
N36 & 7661010104 & $2.62^{+0.16}_{-0.12}$ & $0.010^{+0.001}_{p}$ & $0.645^{+0.009}_{-0.009}$ & $1.195^{+0.019}_{-0.019}$ & 1.65 & 1.837 & 8.22 \\
N38 & 7661010105 & $2.21^{+0.29}_{-0.06}$ & $0.010^{+0.004}_{p}$ & $0.651^{+0.010}_{-0.009}$ & $0.974^{+0.017}_{-0.017}$ & 1.63 & 0.877 & 6.75 \\
N39 & 7661010106 & $2.00^{+0.14}_{p}$ & $0.028^{+0.006}_{-0.002}$ & $0.627^{+0.019}_{-0.017}$ & $0.828^{+0.026}_{-0.027}$ & 1.62 & 0.873 & 5.57 \\
N40 & 7661010107 & $2.00^{+0.18}_{p}$ & $0.023^{+0.006}_{-0.001}$ & $0.667^{+0.014}_{-0.014}$ & $0.770^{+0.018}_{-0.018}$ & 1.62 & 0.617 & 5.45 \\
N42 & 7661010108 & $3.50^{p}_{-0.08}$ & $0.010^{+0.001}_{p}$ & $0.642^{+0.010}_{-0.009}$ & $0.728^{+0.014}_{-0.015}$ & 1.61 & 1.354 & 4.98 \\
N43 & 7661010109 & $3.50^{p}_{-0.19}$ & $0.010^{+0.002}_{p}$ & $0.640^{+0.012}_{-0.011}$ & $0.723^{+0.016}_{-0.017}$ & 1.61 & 1.412 & 4.94 \\
N45 & 7661010110 & $2.71^{+0.73}_{-0.20}$ & $0.010^{+0.010}_{p}$ & $0.675^{+0.022}_{-0.019}$ & $0.470^{+0.018}_{-0.019}$ & 1.57 & 0.885 & 3.36 \\

    \bottomrule
\end{tabular}
\parbox{\textwidth}{\footnotesize \textit{Notes}: The fitting result with Model 3:  {\tt\string tbvarabs(simplcut*kerrbb2)} and the viscosity parameter $\alpha=0.1$. The superscript “p” indicates that the upper and lower limits provided by the fit have reached the boundary constraints.}
\end{table}